\documentclass[aos,MSNbibl,dvips]{arximspdf}
\usepackage{algpseudocode}
\usepackage{algorithm}
\usepackage{algorithmicx}
\usepackage{mathbh}
\usepackage{ifthen}
\usepackage{xargs}

\doi{10.1214/14-AOS1209} 
\volume{42}
\issue{4}
\pubyear{2014}
\firstpage{1483}
\lastpage{1510}

\makeatletter
\newcommand{\nnearrow}{\nearrow}
\newcommand{\ssearrow}{\searrow}
\newcommand{\idotsint}{\int\cdots\int}

\newcommand{\iint}{\int\!\!\int}
\renewcommand{\mid}{|}
\newcommand{\varotimes}{\otimes}
\newtheorem{teo}{Theorem}
\newproclaim{defi}[teo]{Definition}
\newtheorem{lem}[teo]{Lemma}
\newtheorem{prop}[teo]{Proposition}
\newproclaim{rem}[teo]{Remark}
\newproclaim{example}[teo]{Example}

\newcommandx\Aux[2][1=]{\ifthenelse{\equal{#1}{}}
{
\ifthenelse{\equal{#2}{}}{U}{U^{(#2)}}
}
{
\ifthenelse{\equal{#2}{}}{U^{#2}_{#1}}{U^{(#2)}_{#1}}
}
}

\newcommand{\bpi}{\bar{\pi}}
\newcommandx\cY[2][1=]{
\ifthenelse{\equal{#1}{}}{\check{Y}^{(#2)}}{\check{Y}^{(#2)}_{#1}}
}
\newcommandx\cAux[2][1=]{
\ifthenelse{\equal{#1}{}}{\check{U}^{(#2)}}{\check{U}^{(#2)}_{#1}}
}
\newcommandx\cM[2][1=]{
\ifthenelse{\equal{#1}{}}{\check{M}^{(#2)}}{\check{M}^{(#2)}_{#1}}
}
\newcommandx\cZ[2][1=]{
\ifthenelse{\equal{#1}{}}{\check{Z}^{(#2)}}{\check{Z}^{(#2)}_{#1}}
}
\newcommandx{\covar}[3][1=]{\operatorname{Cov}_{#1}(#2,#3)}
\newcommandx{\covardu}[3][1=]{\operatorname{Cov}_{#1}\bigl(#2,#3\bigr)}

\newcommandx{\expect}[2]{
\ifthenelse{\equal{#2}{}}{\mathbb{E}\left[#1\right]}{\mathbb
{E}_{#1}\left[#2\right]}
}

\newcommandx\kH[2][1=]{
\ifthenelse{\equal{#1}{}}
{
\ifthenelse{\equal{#2}{}}{K}{K^{(#2)}}
}
{
\ifthenelse{\equal{#2}{}}{K^{#2}_{#1}}{K^{(#2)}_{#1}}
}
}

\newcommandx\Ltwo[1][1=]{
\ifthenelse{\equal{#1}{}}
{\mathsf{L}^2}
{\mathsf{L}^2(#1)}
}
\newcommandx\Ltwonorm[2][1=]{
\ifthenelse{\equal{#1}{}}
{\Vert #2 \Vert_{\Ltwo}}
{\Vert #2 \Vert_{\Ltwo(#1)}}
}
\newcommandx\M[2][1=]{
\ifthenelse{\equal{#1}{}}
{
\ifthenelse{\equal{#2}{}}{M}{M^{(#2)}}
}
{
\ifthenelse{\equal{#2}{}}{M^{#2}_{#1}}{M^{(#2)}_{#1}}
}
}

\newcommand{\obs}{s_{\mathrm{obs}}}

\newcommandx{\pgeq}[1][1=]{\succcurlyeq}
\newcommandx{\pleq}[1][1=]{\preccurlyeq_{#1}}
\newcommandx{\prior}[2]{
\ifthenelse{\equal{#2}{}}{p_{#1}}{p_{#1}(#2)}
}

\newcommandx\R[2][1=]{
\ifthenelse{\equal{#1}{}}
{
\ifthenelse{\equal{#2}{}}{R}{R^{(#2)}}
}
{
\ifthenelse{\equal{#2}{}}{R^{#2}_{#1}}{R^{(#2)}_{#1}}
}
}
\newcommandx\sequence[3][2=k,3=\mathbb{N}]
{\ifthenelse{\equal{#3}{}}{\ensuremath{\{ #1_{#2}\}}}{\ensuremath{\{
#1_{#2} ; #2 \in #3 \}}}}
\newcommand{\tX}{\hat{X}}
\newcommand{\talpha}{\hat{\alpha}}

\newcommand{\tAuxiv}{\hat{V}}
\newcommand{\tAuxi}{\hat{U}}
\newcommandx\Taux[2][1=]{
\ifthenelse{\equal{#1}{}}
{
\ifthenelse{\equal{#2}{}}{\hat{U}}{\hat{U}^{(#2)}}
}
{
\ifthenelse{\equal{#2}{}}{\hat{U}^{#2}_{#1}}{\hat{U}^{(#2)}_{#1}}
}
}

\newcommandx{\var}[2][1=]{\operatorname{Var}_{#1}\Biggl(#2\Biggr)}
\newcommandx{\varTxt}[2][1=]{\operatorname{Var}_{#1}(#2)}
\newcommandx\X[2][1=]{
\ifthenelse{\equal{#1}{}}
{
\ifthenelse{\equal{#2}{}}{X}{X^{(#2)}}
}
{
\ifthenelse{\equal{#2}{}}{X^{#2}_{#1}}{X^{(#2)}_{#1}}
}
}
\newcommandx\Y[2][1=]{
\ifthenelse{\equal{#1}{}}
{
\ifthenelse{\equal{#2}{}}{Y}{Y^{(#2)}}
}
{
\ifthenelse{\equal{#2}{}}{Y^{#2}_{#1}}{Y^{(#2)}_{#1}}
}
}
\newcommand{\tY}{\hat{Y}}
\newcommandx\Z[2][1=]{
\ifthenelse{\equal{#1}{}}
{
\ifthenelse{\equal{#2}{}}{Z}{Z^{(#2)}}
}
{
\ifthenelse{\equal{#2}{}}{Z^{#2}_{#1}}{Z^{(#2)}_{#1}}
}
}
\makeatother

\begin{document}
\begin{frontmatter}

\title{Comparison of asymptotic variances of inhomogeneous Markov
chains with application to Markov chain Monte Carlo methods}
\runtitle{Comparison of variances of inhomogeneous Markov chains\quad }

\begin{aug}
\author[a]{\fnms{Florian} \snm{Maire}\corref{}\ead[label=e1]{florian.maire@it-sudparis.eu}\thanksref{tt1}},
\author[a]{\fnms{Randal} \snm{Douc}\ead[label=e2]{randal.douc@it-sudparis.eu}\thanksref{tt1}}
\and
\author[b]{\fnms{Jimmy} \snm{Olsson}\ead[label=e3]{jimmyol@kth.se}\thanksref{tt2}}
\runauthor{F. Maire, R. Douc and J. Olsson}
\affiliation{Institut T\'el\'ecom/T\'el\'ecom SudParis and CNRS UMR 5157
SAMOVAR,\\
Institut T\'el\'ecom/T\'el\'ecom SudParis and CNRS UMR 5157 SAMOVAR,\\
and
KTH Royal Institute of Technology}
\address[a]{F. Maire\\
R. Douc\\
Telecom SudParis\\
9 rue Charles Fourier\\
91011 Evry\\
France\\
\printead{e1}\\
\phantom{E-mail: }\printead*{e2}} 
\address[b]{J. Olsson\\
KTH Royal Institute of Technology\\
SE-100 44 Stockholm\\
Sweden\\
\printead{e3}}
\end{aug}
\thankstext{tt1}{Supported in part by the ONERA, the French Aerospace
Lab and the DGA, the French Procurement Agency.}
\thankstext{tt2}{Supported by the Swedish Research Council, Grant 2011-5577.}

\received{\smonth{7} \syear{2013}}
\revised{\smonth{12} \syear{2013}}

%
\begin{abstract}
In this paper, we study the asymptotic variance of sample path averages
for inhomogeneous Markov chains that
evolve alternatingly according to two different $\pi$-reversible
Markov transition
kernels\break  $P$ and $Q$. More specifically, our main result allows us to
compare directly the asymptotic
variances of two inhomogeneous Markov chains associated with different
kernels $P_i$ and $Q_i$, $i \in\{0, 1\}$,
as soon as the kernels of each pair $(P_0, P_1)$ and $( Q_0, Q_1)$ can
be ordered in the sense of lag-one
autocovariance. As an important application, we use this result for
comparing different data-augmentation-type
Metropolis--Hastings algorithms. In particular, we compare some
pseudo-marginal\break  \mbox{algorithms} and propose a novel
exact algorithm, referred to as the \textit{random refreshment}
algorithm, which is more efficient, in terms of asymptotic
variance, than the Grouped Independence Metropolis--Hastings algorithm
and has a computational complexity that
does not exceed that of the Monte Carlo Within Metropolis algorithm.
\end{abstract}

%
\end{frontmatter}

\section{Introduction} \label{secintroduction}\emph{Markov chain Monte
Carlo} (MCMC) \emph{methods} allow samples from virtually any target
distribution $\pi$, known up to a normalizing constant, to be
generated. In particular, the celebrated \emph{Metropolis--Hastings
algorithm} (introduced in \cite{metropolis1953} and \cite
{hastings1970}) simulates a Markov chain evolving according to a \mbox
{$\pi
$-}reversible Markov transition kernel by first generating, using some
instrumental kernel, a candidate and then accepting or rejecting the
same with a probability adjusted to satisfy the detailed balance
condition~\cite{tierneynote}. When choosing between several
Metropolis--Hastings algorithms, it is desirable to be able to compare
the efficiencies, in terms of the asymptotic variance of sample path
averages, of different $\pi$-reversible Markov chains. Despite the
practical importance of this question, only a few results in this
direction exist the literature. Peskun \cite{peskunoptimum}
defined a partial ordering for finite state space Markov chains, where
one transition kernel has a higher order than another if the former
dominates the latter on the off-diagonal (see Definition~\ref
{defiPeskunordering}). This ordering was extended later by Tierney
\cite{tierneynote} to general state space Markov chains and another
even more general ordering, the covariance ordering, was proposed in
\cite{miraordering}. In general, it holds that if a homogeneous $\pi
$-reversible Markov transition kernel is greater than another according
to one of these orderings, then the asymptotic variance of sample path
averages for a Markov chain evolving according to the former is smaller
for all 
square integrable (with respect to $\pi$) target functions.

We provide an extension of this result to inhomogeneous Markov chains
that evolve alternatingly according to two different $\pi$-reversible
Markov transition kernels. To the best of our knowledge, this is the
first work dealing with systematic comparison of asymptotic variances
of inhomogeneous Markov chains. The approach is linked with the
operator theory for Markov chains but does not make use of any spectral
representation.
After some preliminaries (Section~\ref{secpreliminaries}), our main
result, Theorem~\ref{teomainResult}, is stated in Section~\ref
{secmain}. In Section~\ref{secappl}, we apply Theorem~\ref
{teomainResult} in the context of MCMC algorithms by comparing the
efficiency, in terms of asymptotic variance, of some existing
data-augmentation-type algorithms. Moreover, we propose a novel
pseudo-marginal algorithm (in the sense of \cite{andrieupseudo}),
referred to as the
\textit{random refreshment} algorithm, which---on the contrary to the
pseudo-marginal version of the \emph{Monte Carlo Within Metropolis}
(MCWM) algorithm---turns out to be exact and more efficient than the
pseudo-marginal version of the \emph{Grouped Independence
Metropolis--Hastings} (GIMH) algorithm. Here, the analysis is again
driven by Theorem~\ref{teomainResult}.
The proof of Theorem~\ref{teomainResult} is given in Section~\ref
{secproofmain} and some technical lemmas are postponed to
Appendix~\ref{app}. Finally, Appendix~\ref{secappB} relates some
existing MCMC algorithms to the framework considered in this paper.

\section{Preliminaries} \label{secpreliminaries}We denote by $\mathbb{N}:=\{0, 1, 2, \ldots\}$ and $\mathbb{N}^{\ast}:=\{1, 2, \ldots\}$ the
sets of nonnegative and positive integers, respectively. In the
following, all random variables are assumed to be defined on a common
probability space $(\Omega, {\mathcal F}, \mathbb{P})$. Let $(\mathsf{X},
\mathcal{X})$ be a measurable space; then we denote by $\mathcal
{M}(\mathcal{X})$ and
$\mathcal{F}(\mathcal{X})$ the spaces of positive measures and measurable
functions on $(\mathsf{X}, \mathcal{X})$, respectively. The Lebesgue
integral of $f
\in\mathcal{F}(\mathcal{X})$ over $\mathsf{X}$ with respect to the
measure $\mu\in
\mathcal{M}(\mathcal{X})$ is, when well-defined, denoted by $\mu f:=\int f(x)
\mu(\mathrm{d}x)$. Recall that a \emph{Markov transition kernel} $P$
on $(\mathsf{X}, \mathcal{X})$ is a mapping $P\dvtx\mathsf{X}\times
\mathcal{X}
\rightarrow
[0, 1]$ such that:
\begin{itemize}
\item for all $\mathsf{A} \in\mathcal{X}$, $\mathsf{X}\ni x \mapsto
P(x, \mathsf{A})$ is a measurable function,
\item for all $x \in\mathsf{X}$, $\mathcal{X}\ni\mathsf{A} \mapsto
P(x, \mathsf{A})$ is a probability measure.
\end{itemize}
A kernel $P$ induces two integral operators, one acting on $\mathcal
{M}(\mathcal{X})$ and the other on $\mathcal{F}(\mathcal{X})$; more
specifically, for $\mu
\in\mathcal{M}(\mathcal{X})$ and $f \in\mathcal{F}(\mathsf{X})$,
we define the measure
\[
\mu P\dvtx\mathcal{X}\ni\mathsf{A} \mapsto\int P(x, \mathsf{A}) \mu(\mathrm{d}x)
\]
and the measurable function
\[
Pf \dvtx\mathsf{X}\ni x \mapsto\int f \bigl(x' \bigr) P \bigl(x,
\mathrm{d}x' \bigr).
\]
Moreover, the \emph{composition} (or \emph{product}) of two kernels
$P$ and $Q$ on $(\mathsf{X}, \mathcal{X})$ is the kernel defined by
\[
PQ\dvtx\mathsf{X}\times\mathcal{X}\ni(x, \mathsf{A}) \mapsto\int Q
\bigl(x', \mathsf{A} \bigr) P \bigl(x,\mathrm{d}x'
\bigr).
\]
We will from now on fix a distinguished probability measure $\pi$ on
$(\mathsf{X}, \mathcal{X})$. Given~$\pi$, we denote by $\Ltwo[\pi]:=
\{ f
\in\mathcal{F}(\mathcal{X}) \dvtx\pi f^2 < \infty\}$ the space of square
integrable functions with respect to $\pi$ and furnish the same with
the scalar product
\[
\langle f,  g \rangle:=\int f(x) g(x) \pi(\mathrm{d}x) \qquad \bigl(f \in
\Ltwo[ \pi], g \in\Ltwo[\pi] \bigr)
\]
and the associated norm
\[
\Ltwonorm{f}:= \bigl( \pi f^2 \bigr)^{1/2} \qquad \bigl(f
\in\Ltwo[\pi] \bigr).
\]
Here, we have expunged the measure $\pi$ from the notation for
brevity. If $P$ is a Markov\vspace*{1.5pt} kernel on $(\mathsf{X}, \mathcal{X})$ admitting
$\pi$
as an invariant distribution, then the mapping $f \mapsto Pf$ defines
an operator on $\Ltwo[\pi]$, and by Jensen's inequality it holds that
%
\begin{equation}
\label{eqmajoNormP} \| P \|:=\sup_{f \in\Ltwo[\pi] \dvtx
\Ltwonorm{f} \leq1} \Ltwonorm{Pf} \leq1.
\end{equation}
Recall that a kernel $P$ is \emph{$\pi$-reversible} if and only if
the detailed balance relation
\[
\pi(\mathrm{d}x) P \bigl(x,\mathrm{d}x' \bigr) = \pi \bigl(
\mathrm{d} x' \bigr) P \bigl(x',\mathrm{d}x \bigr)
\]
holds. If the Markov kernel $P$ is $\pi$-reversible, then $f \mapsto
P f$ defines a self-adjoint operator on $\Ltwo[\pi]$, that is, for
all $f$ and $g$ belonging to $\Ltwo[\pi]$,
%
\begin{equation}
\label{eqpiReversibility} \langle f,  Pg \rangle = \langle Pf,  g \rangle.
\end{equation}

The following off-diagonal ordering of Markov transition kernels on a
common state space was, in the case of Markov chains in a finite state
space, proposed in~\cite{peskunoptimum}. The ordering was extended
later in \cite{tierneynote} to the case of Markov chains in general state space.
%
\begin{defi} \label{defiPeskunordering}
Let $P_0$ and $P_1$ be Markov transition kernels on $(\mathsf{X},
\mathcal{X})$
with invariant distribution $\pi$. We say that \emph{$P_1$ dominates
$P_0$ on the off-diagonal}, denoted $P_1 \succeq P_0$, if for all
$\mathsf{A} \in\mathcal{X}$ and $\pi$-a.s. all $x \in\mathsf{X}$,
\[
P_1 \bigl(x, \mathsf{A} \setminus\{ x \} \bigr) \geq
P_0 \bigl(x, \mathsf{A} \setminus\{ x \} \bigr).
\]
\end{defi}
The previous ordering allows the asymptotic efficiencies of different reversible kernels to be compared.
 More specifically, the following seminal
result was established in \cite{peskunoptimum}, Theorem 2.1.1, for
Markov chains in discrete state space and extended later in \cite
{tierneynote}, Theorem 4, to Markov chains in general state space.
%
\begin{teo} \label{teoefficiencyOrdering}
Let $P_0$ and $P_1$ be two $\pi$-reversible kernels on $(\mathsf{X},
\mathcal{X}
)$. If \mbox{$P_1 \succeq P_0$}, then for a.s. all $f \in\Ltwo[\pi]$,
\[
v(f, P_1) \leq v(f, P_0),
\]
where we have defined, for a Markov chain $\sequence{\X{}}[k][\mathbb{N}]$
with $\pi$-reversible transition kernel $P$ and initial distribution
$\pi$,
%
\begin{equation}
\label{eqdefvarasymp} v(f,P):=\lim_{n \rightarrow\infty} \frac{1}{n} \var{ \sum
_{k=0}^{n-1}f(X_k)}.
\end{equation}
\end{teo}
Note\vspace*{1pt} that according to \cite{tierneynote}, if $\sequence{X}$ is a
$\pi
$-reversible Markov chain and $f \in\Ltwo[\pi]$, then $\lim_{n
\rightarrow\infty} n^{-1} \varTxt{\sum_{k = 0}^{n - 1}f(X_k)}$ is
guaranteed to exist (but may be infinite). Nevertheless, the ordering
in question does not allow Markov kernels lacking probability mass on
the diagonal, that is, kernels $P$ satisfying \mbox{$P(x, \{ x \}) = 0$} for
all $x \in\mathsf{X}$, to be compared. This is in particular the case for
Gibbs samplers in general state space. To overcome this limitation, one
may consider instead the following covariance ordering based on lag-one
autocovariances.
%
\begin{defi} \label{deficovarordering}
Let $P_0$ and $P_1$ be Markov transition kernels on $(\mathsf{X},
\mathcal{X})$
with invariant distribution $\pi$. We say that \emph{$P_1$ dominates
$P_0$ in the covariance ordering}, denoted $P_1 \pgeq[1] P_0$, if for
all $f \in\Ltwo[\pi]$,
\[
\langle f,  P_1f \rangle \leq \langle f,  P_0f \rangle.
\]
\end{defi}
The covariance ordering, which was introduced implicitly in \cite
{tierneynote}, page~5, and formalized in \cite{miraordering}, is an
extension of the off-diagonal ordering since according to~\cite
{tierneynote}, Lemma~3, $P_1\succeq P_0$ implies $P_1 \pgeq[1] P_0$.
Moreover, it turns out that for reversible kernels, $P_1 \pgeq[1]P_0$
implies $v(f,P_0) \geq v(f,P_1)$ (see the proof of \cite{tierneynote},
Theorem~4).

All these results concern homogeneous Markov chains, whereas many MCMC
algorithms such as the Gibbs or the Metropolis-within-Gibbs samplers
use several kernels, for example, $P$ and $ Q $ in the case of two
kernels \cite{RobertMCMC}. A natural idea would then be to apply
Theorem~\ref{teoefficiencyOrdering} to the homogeneous Markov chain
having the block kernel $P Q$ as transition kernel; however, even when
the kernels $P$ and $ Q$ are both $\pi$-reversible, the product $P Q$
of the same is usually not $\pi$-reversible, except in the particular
case when $P$ and $ Q$ commute, that is, $P Q= Q P$. Thus, Theorem~\ref
{teoefficiencyOrdering} cannot in general be applied directly in this case.

\section{Main assumptions and results} \label{secmain}In the
following, let $P_i$ and $ Q_i$, $i \in\{0, 1\}$, be Markov transition
kernels on $(\mathsf{X}, \mathcal{X})$. Define $\sequence{\X{0}}$ and
$\sequence{\X
{1}}$ as the Markov chains evolving as follows:
%
\begin{equation}
\label{eqeq1Markov} \X[0]{i} \stackrel{P_i} {\longrightarrow} \X[1]{i}
\stackrel{ Q_i} {\longrightarrow} \X[2]{i} \stackrel{P_i}
{\longrightarrow} \X[3]{i} \stackrel{ Q_i} {\longrightarrow} \cdots.
\end{equation}
This means that for all $k \in\mathbb{N}$, $i \in\{0, 1\}$ and
$\mathsf{A}
\in\mathcal{X}$:
\begin{itemize}
\item$\mathbb{P} ( \X[2 k+1]{i} \in\mathsf{A} \mid\mathcal{F}_{2
k}^{(i)} ) = P_i(\X[2 k]{i}, \mathsf{A})$,\vspace*{2pt}
\item$\mathbb{P} ( \X[2 k+2]{i} \in\mathsf{A} \mid\mathcal{F}_{2
k+1}^{(i)} ) = Q_i(\X[2 k+1]{i},\mathsf{A})$,
\end{itemize}
where $\mathcal{F}_{n}^{(i)}:=\sigma(\X[0]{i},\ldots, \X[n]{i})$,
$n \in\mathbb{N}$. We impose the following assumption:
{\renewcommand{\theequation}{A1}
\begin{eqnarray}\label{assPeskunorder}
\hspace*{-100pt}&& \mbox{\phantom{i}(i) $P_i$ and $ Q_i$, $i \in\{0, 1\}$, are $\pi$-reversible,}
\nonumber\\[-8pt] \hspace*{-100pt}\\[-8pt]
\hspace*{-100pt}&&\mbox{(ii) $P_1 \pgeq[1] P_0$ and $ Q_1 \pgeq[1] Q_0$.}\nonumber
\end{eqnarray}}%
As mentioned above, $P_1\succeq P_0$ implies $P_1\pgeq[1]P_0$; thus, in
practice, a sufficient condition for \hyperref[assPeskunorder]{(A1)}(ii) is that $P_1 \succeq P_0$ and $ Q_1
\succeq Q_0$.

\setcounter{equation}{4}
%
\begin{teo} \label{teomainResult}
Assume that $P_i$ and $ Q_i$, $i \in\{0,1\}$, satisfy \textup{\hyperref[assPeskunorder]{(A1)}} and let $\sequence{\X{i}}$, $i \in\{0,1\}$,
be Markov chains evolving as in (\ref{eqeq1Markov}) with initial
distribution $\pi$.
Then for all $f \in\Ltwo[\pi]$ such that for $i \in\{0,1\}$,
%
\begin{equation}
\label{eqassumpFuncThm} \sum_{k = 1}^{\infty} \bigl( \bigl|
\covardu{f \bigl(\X[0]{i} \bigr)} {f \bigl(\X[k]{i} \bigr)}\bigr|+\bigl|\covardu {f \bigl(
\X[1]{i} \bigr)} {f \bigl(\X[k+1]{i} \bigr)}\bigr| \bigr) < \infty,
\end{equation}
it holds that
%
\begin{equation}
\label{eqmainresult} v_{1}(f) \leq v_{0}(f),
\end{equation}
where
%
\begin{equation}
\label{eqmainresultlimits} v_{i}(f):=\lim_{n \to\infty}
\frac{1}{n} \var{\sum_{k=0}^{n-1}f
\bigl( \X[k]{i} \bigr)} \qquad \bigl(i \in\{0, 1\} \bigr).
\end{equation}
\end{teo}

%
\begin{rem}
At present, we have not been able to extend the arguments of our
current proof of Theorem~\ref{teomainResult} (see Section~\ref
{secproofmain}) to inhomogeneous Markov chains evolving alternatingly
according to \emph{more} than two different kernels. On the other hand,
we have not been able to find a counterexample rejecting the hypothesis
that a similar result would hold true also in that case. We leave this
as an open problem.
\end{rem}

%
\begin{rem} \label{remcounterexsummability}
Condition (\ref{eqassumpFuncThm}) is \emph{not} a necessary
condition for (\ref{eqmainresult}); indeed, letting $\mathsf{X} = \{-1,
1\}$, $\pi(\mathrm{d}x') = P_0(x, \mathrm{d}x') = (\delta_1(\mathrm
{d}x') + \delta
_{-1}(\mathrm{d}x'))/2$, $ Q_1 = Q_0 = P_1$, where, as in \cite
{haggstrom2007variance}, Example~5, $P_1(x, \mathrm{d}x') = \delta
_{-x}(\mathrm{d}
x')$, provides a straightforward counterexample.
\end{rem}

When verifying if a given $f$ satisfies the condition (\ref{eqassumpFuncThm})
it may be convenient to consider the homogeneous Markov chains $\{
X_{2k}; k \in\mathbb{N}\}$ or $\{X_{2k+1}; k \in\mathbb{N}\}$ or
even $\{(X_{2k}, X_{2k+1}); k \in\mathbb{N}\}$. Typically, none of these chains
are $\pi$-reversible. Nevertheless, $\pi$-reversibility is not needed
for checking conditions of type (\ref{eqassumpFuncThm}), which can be
established using upper bounds on the \emph{$V$-norm} between the
distribution given by the $n$th iterate of a homogeneous
kernel and its stationary distribution. This will be developed in the
following section.


\subsection{Sufficient conditions for the absolute summability
assumption (\texorpdfstring{\protect\ref{eqassumpFuncThm}}{5})}

For any measurable real-valued function $f$ on $(\mathsf{X},\mathcal
{X})$, define
the \emph{$V$-norm} of the \emph{function} $f$ by
\[
|f|_V:=\sup_{x \in\mathsf{X}} \frac{|f(x)|}{V(x)}.
\]
Moreover, let $\xi$ be a finite signed measure on $(\mathsf
{X},\mathcal{X})$. Then
by the Jordan decomposition theorem there exists a unique pair of
positive, finite and singular measures $\xi_+$ and $\xi_-$ on
$(\mathsf{X},\mathcal{X})$ such that $\xi= \xi_+ - \xi_-$. The pair $\xi_\pm$ is
referred to as
the \emph{Jordan decomposition} of the signed measure $\xi$. The finite
measure $|\xi|:=\xi_+ + \xi_-$ is
called the \emph{total variation} of $\xi$. Let $V$ be a nonnegative
function taking values in $[1,\infty)$; then the \emph{$V$-norm} of the
\emph{signed measure} $\xi$ is defined by
\[
\ensuremath{\Vert\xi\Vert_{V}}:=| \xi|(V) = \sup
_{f \dvtx|f|_V
\leq1} \xi f.
\]

%
\begin{defi} \label{defVgeomerg}
A Markov kernel $P$ on $(\mathsf{X},\mathcal{X})$ is \emph{$V$-geometrically
ergodic} if it admits a unique invariant distribution $\pi$ and there
exists a measurable function $V\dvtx\mathsf{X}\to[1,\infty)$ satisfying
$\pi V
< \infty$ and such that the following hold:
\begin{longlist}[(a)]
\item[(a)] There exist constants $(C,\rho) \in\mathbb{R}^+\times(0,1)$ such
that for all $x \in\mathsf{X}$ and all $n \in\mathbb{N}$,
%
\begin{equation}
\label{eqVgeom} \ensuremath{\bigl\Vert P^n(x,\cdot) - \pi
\bigr\Vert_{V}} \leq C \rho^n V(x).
\end{equation}
\item[(b)] There exist constants $(b, \lambda) \in\mathbb{R}^+ \times(0,1)$
such that $PV \leq\lambda V + b$.\vadjust{\goodbreak}
\end{longlist}
\end{defi}

%
\begin{rem}
\cite{hairer2011yet}, Theorem~1.2, provides sufficient conditions, in
terms of drift towards a \emph{small set}, for (a) in Definition~\ref
{defVgeomerg} to hold; see also \cite{roberts2004general}, Fact~10,
for necessary and sufficient conditions under the assumption of
aperiodicity and irreducibility. Moreover, the coming developments
require only the bound (\ref{eqVgeom}) to hold $\pi$-a.s.
\end{rem}

We have now all necessary tools for giving sufficient conditions that
imply the absolute summability assumption (\ref{eqassumpFuncThm}). Let
the chain $\sequence{X}$ evolve according to
%
\begin{equation}
\label{eqeq1Markovbis} \X[0]{} \stackrel{P} {\longrightarrow} \X [1]{} \stackrel{Q} {
\longrightarrow} \X[2]{} \stackrel{P} {\longrightarrow} \X[3]{} \stackrel{Q} {
\longrightarrow} \cdots
\end{equation}
with $\X[0]{} \sim\pi$, for some Markov kernels $P$ and $Q$.
%
\begin{prop} \label{propaltcondition}
If the Markov kernel $P Q$ is $V$-geometrically ergodic, then for all
functions $f$ such that $|f|_{V^{1/2}} < \infty$ and $|Pf|_{V^{1/2}} <
\infty$,
\[
\sum_{k = 1}^{\infty} \bigl( \bigl|\covardu{f \bigl(
\X[0]{} \bigr)} {f \bigl(\X[k]{} \bigr)}\bigr|+\bigl|\covardu{f \bigl(\X [1]{} \bigr)} {f
\bigl( \X[k+1]{} \bigr)}\bigr| \bigr) < \infty,
\]
where $\{\X[k]{}; k \in\mathbb{N}\}$ evolves as in (\ref{eqeq1Markovbis}).
\end{prop}
The proof of Proposition~\ref{propaltcondition} is found in
Appendix~\ref{appproofpropaltcondition}.

\section{Application to data-augmentation-type algorithms} \label
{secappl}Before considering some applications of Theorem~\ref
{teomainResult}, we recall the following proposition, describing how
to obtain a $\pi$-reversible Markov chain using some instrumental
kernel $K$. Although this result is fundamental in the
Metropolis--Hastings literature (see, e.g., \cite{roberts2004general,RobertMCMC,gilks1996markov} and the references therein),
it is restated here as it will be used in various situations in the
sequel [especially when there is no fixed reference measure dominating
all the distributions $\{K(x, \cdot); x \in\mathsf{X}\}$].

\begin{prop}
\label{propderiveeRadon}
Let $K$ be a Markov transition kernel on $\mathsf{X}\times\mathcal
{X}$ and
$\pi
$ a probability measure on $(\mathsf{X}, \mathcal{X})$. Define the probability
measures $\mu(\mathrm{d}x \times\mathrm{d}x'):=\pi(\mathrm{d}x)
K(x,\mathrm{d}
x')$ and $\nu(\mathrm{d}x \times\mathrm{d}x'):=\pi(\mathrm{d}x') K(x',
\mathrm{d}x)$. Assume that the measures $\nu$ and $\mu$ are
equivalent and
such that for $\mu$-a.s. all $(x, x') \in\mathsf{X}^2$,
%
\begin{equation}
\label{eqcondradon} 0 < \frac{\mathrm{d}\nu}{\mathrm{d}\mu} \bigl(x, x' \bigr) < \infty,
\end{equation}
where\vspace*{1pt} $\frac{\mathrm{d}\nu}{\mathrm{d}\mu}$ denotes the
Radon--Nikodym derivative.
Then the Markov kernel $P(x,\mathrm{d}x'):=K(x,\mathrm{d}x') \alpha(x,
x') + \delta_{x}(\mathrm{d}x') \beta(x)$, where
\[
\alpha \bigl(x, x' \bigr):=1\wedge\frac{\mathrm{d}\nu}{\mathrm
{d}\mu
} \bigl(x,
x' \bigr)\quad\mbox{and}\quad\beta(x):=1 - \int K \bigl(x,
\mathrm{d}x' \bigr) \alpha \bigl(x, x' \bigr),
\]
is $\pi$-reversible.
\end{prop}
A natural application of Theorem~\ref{teomainResult} consists in using
the result for comparing different data-augmentation-type algorithms.
In the following, we wish to target a probability distribution $\pi
^{\ast}$
defined on $(\mathsf{Y},\mathcal{Y})$ using a sequence $\sequence{\Y
{}}[k][\mathbb{N}]$
of \mbox{$\mathsf{Y}$-}valued random variables. To this aim, Tanner and
Wong \cite{tannercalculation} suggest writing $\pi^{\ast}$ as the
marginal of some distribution $\pi$ defined on the product space
\mbox{$(\mathsf{Y}\times\mathsf{U}, \mathcal{Y}\varotimes\mathcal{U})$}
in the sense that
$\pi(\mathrm{d}y \times\mathrm{d}u) = \pi^{\ast}(\mathrm{d}y)
R(y,\mathrm{d}u)$,
where $R$ is some Markov transition kernel on $\mathsf{Y}\times
\mathcal{U}$. In
most cases, the marginal $\pi^{\ast}$ is of sole interest, while the
component $u$ is introduced for convenience as a means of coping with
analytic intractability of the marginal. (It could also be the case
that the marginal $\pi^{\ast}$ is too computationally expensive to
evaluate.) A first solution consists in letting $\sequence{\Y
{}}[k][\mathbb{N}]$ be the first-component process $\sequence{\Y
{1}}[k][\mathbb{N}
]$ of the $\pi$-reversible Markov chain $\{(\Y[k]{1}, \Aux[k]{1}); k
\in\mathbb{N}\}$ defined as follows. Let $S$ and $T$ be instrumental Markov
transition kernels on $\mathsf{Y}\times\mathsf{U}\times\mathcal
{Y}$ and $\mathsf{Y}
\times
\mathsf{U}\times\mathsf{Y}\times\mathcal{U}$, respectively, and
define a
transition of the chain $\{(\Y[k]{1}, \Aux[k]{1}); k \in\mathbb
{N}\}$ by
Algorithm~\ref{algalg1}.
\renewcommand{\algorithmicrequire}{Given}
\renewcommand{\algorithmicensure}{\textbf{Output:}}
\begin{algorithm}[t]
\begin{algorithmic}[5] \caption{The \textit{freeze} algorithm} \label
{algalg1}
\Require$(\Y[k]{1}, \Aux[k]{1}) = (y, u)$:
\begin{longlist}[(iii)]
\item[(i)] draw $\tY\sim S(y, u; \cdot)$ and call the outcome $\hat{y}$
(abbr. $\leadsto\hat{y}$),
\item[(ii)] draw $\tAuxi\sim T(y, u, \hat{y}; \cdot) \leadsto\hat{u}$,
\item[(iii)] set
%
\begin{eqnarray}\label{eqacceptMetropolis}
&& \bigl(\Y[k+1]{1}, \Aux[k+1]{1} \bigr)
\nonumber\\[10pt]\\[-22pt]
&&\qquad \gets
\cases{\displaystyle (\hat{y}, \hat{u}), &\quad with probability $\displaystyle\alpha (y, u, \hat{y}, \hat{u})$
\vspace*{5pt}\cr
&\qquad\quad $\displaystyle:=1 \wedge\frac{ \pi^{\ast}(\hat{y}) r(\hat{y}, \hat{u})s(\hat{y}, \hat{u}; y) t(\hat{y}, \hat{u}, y; u)}{\pi^{\ast}(y)
r(y, u) s(y, u; \hat{y}) t(y,u, \hat{y}; \hat{u})}$,
\vspace*{5pt}\cr
(y, u), &\quad otherwise.}\nonumber
\end{eqnarray}
\end{longlist}
\end{algorithmic}
\end{algorithm}
%
%
\begin{rem} \label{remgeneralradon}
In the expression (\ref{eqacceptMetropolis}) of $\alpha,$ we assume
implicitly that the families $\{S(y, u; \cdot); (y, u) \in
\mathsf{Y}\times\mathsf{U}\}$ and $\{T(y, u, \hat{y}; \cdot); (y, u,
\hat{y}
) \in\mathsf{Y}\times\mathsf{U}\times\mathsf{Y}\}$ of probability
measures are
dominated by a fixed nonnegative measure and we denote by $s$ and $t$
the corresponding transition kernel densities, respectively. In some
cases (see, e.g., \cite{nichollscoupled}) it may, however, happen
(typically when some Dirac mass is involved) that these kernels are not
dominated by a nonnegative measure; nevertheless,
Algorithm~\ref{algalg1} as well as Algorithm~\ref{algalg2} defined
below remain valid provided that the ratio in $\alpha$ is replaced by
the corresponding Radon--Nikodym derivative
$
\frac{\mathrm{d}\nu}{\mathrm{d}\mu}(y, u, \hat{y}, \hat{u}),
$
where in this case,
\begin{eqnarray*}
\mu(\mathrm{d}y\times\mathrm{d}u\times\mathrm {d}
\hat{y}\times\mathrm{d} \hat{u}) &:=&\pi(\mathrm{d}y) R(y, \mathrm{d}u) S(y, u;
\mathrm{d}\hat{y}) T(y, u, \hat{y}; \mathrm{d}\hat{u}),
\\
\nu(\mathrm{d}y\times\mathrm{d}u\times\mathrm{d}\hat{y}\times \mathrm{d}\hat{u})
&:=&\pi(\mathrm{d}\hat{y}) R( \hat{y}, \mathrm{d}\hat{u}) S(\hat{y}, \hat{u};
\mathrm{d}y) T(\hat{y}, \hat{u}, y; \mathrm{d}u).
\end{eqnarray*}
\end{rem}

By applying Proposition~\ref{propderiveeRadon}, we deduce that the
output $\{(\Y[k]{1}, \Aux[k]{1});\break  k\in\mathbb{N}\}$ is a $\pi
$-reversible Markov chain. As a consequence, the sequence $\sequence
{\Y
{1}}[k][\mathbb{N}]$ targets, although it is not itself a Markov
chain, the
marginal distribution $\pi^{\ast}$. Note that the method requires the
product $\pi^{\ast}(y) r(y, u) s(y, u;\break  \hat{y}) t(y, u, \hat{y};
\hat{u})$ to be known at least up to a multiplicative constant to
guarantee the computability of the acceptance probability $\alpha$ in
(\ref{eqacceptMetropolis}).

%
\begin{example}[(Grouped Independence Metropolis--Hastings)]\label{exgimh}
The Group\-ed Independence Metropolis--Hastings (GIMH) algorithm (see
\cite{beaumontestimation,andrieupseudo}) is used in situations where
$\pi^{\ast}$ is analytically intractable. In this algorithm, the quantity
$\pi^{\ast}(y)$ is in the acceptance probability replaced by an importance
sampling estimate
%
\begin{equation}
\label{eqGIMHMCest} \pi^*_N(y):=\frac{1}{N} \sum
_{\ell= 1}^N \frac{\bpi(y, v_\ell
)}{q_y(v_\ell)},
\end{equation}
where $\bpi(y, v)$ is the density of some augmented target distribution
$\bpi(\mathrm{d}y \times\mathrm{d}v)$ defined on the product space
$(\mathsf{Y}
\times
\mathsf{V}, \mathcal{Y}\varotimes\mathcal{V})$, known up to a
normalizing constant and
allowing $\pi^{\ast}$ as marginal distribution, and $\{ v_1, \ldots, v_N
\}
$ are i.i.d. draws from the proposal $q_y$. Denoting by $s(y, \cdot)$
the density used for proposing new candidates $\hat{y}$, one obtains the
acceptance probability ratio
\[
\frac{\pi^*_N(\hat{y}) s(\hat{y}, y)}{\pi^*_N(y) s(y, \hat
{y})} = \frac{\pi^{\ast}
(\hat{y}) r(\hat{y}, \hat{u}) s(\hat{y}, y) t(y, u)}{\pi^{\ast}(y
) r(y, u)
s(y, \hat{y}) t(\hat{y}, \hat{u})},
\]
where $u:=(v_1, \ldots, v_N)$ and
\begin{eqnarray*}
\pi^{\ast}(y) r(y, u) &=& \frac{1}{N} \sum
_{\ell= 1}^N \biggl( \bpi(y, v_\ell)
\prod_{m \neq\ell} q_y(v_m) \biggr),
\\
t(y, u) &=& \prod_{\ell= 1}^N
q_y(v_\ell).
\end{eqnarray*}
Consequently, the GIMH algorithm can be perfectly cast into the
framework of the freeze algorithm, with the auxiliary variable $U$
playing the role of the \mbox{$N$-}dimensional Monte Carlo sample and $\mathsf{U}
=\mathsf{V}^n$.
\end{example}

In the following, we use Theorem~\ref{teomainResult} for comparing the
performance of Algorithm~\ref{algalg1} to that of different
modifications of the same obtained in the cases where:
\begin{longlist}[(II)]
\item[(I)] simulating $R$-transitions is feasible,
\item[(II)] simulating $R$-transitions is infeasible.
\end{longlist}


%
\begin{algorithm}[t]
\begin{algorithmic}[6]
\caption{The \textit{systematic refreshment} algorithm} \label{algalg2}
\Require$\Y[k]{2} = y$:
\begin{longlist}[(iii)]
\item[(i)] draw $U\sim R(y, \cdot) \leadsto u$,
\item[(ii)] draw $\tY\sim S(y,u; \cdot) \leadsto\hat{y}$,
\item[(iii)] draw $\tAuxi\sim T(y,u,\hat{y}; \cdot) \leadsto\hat{u}$,\vspace*{2pt}
\item[(iv)] set
$
 \Y[k+1]{2} \gets
\cases{
\hat{y}, &\quad  with probability $\alpha(y, u, \hat{y}, \hat{u})$ [defined in (\ref{eqacceptMetropolis})],
\vspace*{2pt}\cr
y, &\quad otherwise.}
$
\end{longlist}
\end{algorithmic}
\end{algorithm}

\subsection*{Case \textup{I}: Simulating $R$-transitions is feasible}

In this case, an alternative to Algorithm~\ref{algalg1} consists in
letting $\sequence{\Y{}}[k][\mathbb{N}]$ be the sequence $\sequence
{\Y
{2}}[k][\mathbb{N}]$ generated through Algorithm~\ref{algalg2}.
Note that Algorithm~\ref{algalg2} ``refreshes,'' in step~(i),
systematically the second component of the Markov chain, which
advocates Algorithm~\ref{algalg2} to have better mixing properties
than Algorithm~\ref{algalg1}. The main task of the present section is
to establish rigorously this heuristics. The output $\sequence{\Y{2}}$
of Algorithm~\ref{algalg2} is, on the contrary to $\sequence{\Y{1}}$,
a Markov chain. It is not a classical Metropolis--Hastings Markov chain
due to the auxiliary variables $\Aux{}$ and $\Taux{}$ that appear
explicitly in the acceptance probability. However, as established in
the following proposition, whose proof\vspace*{1.5pt} is found in Appendix~\ref{appproofrevsystrefresh}, the $\pi$-reversibility of $\{(\Y
[k]{1},\Aux[k]{1}); k\in\mathbb{N}\}$ implies $\pi^{\ast
}$-reversibility of
$\sequence{\Y{2}}$.

%
\begin{prop} \label{propinduces_rever}
The sequence $\sequence{\Y{2}}$ generated in Algorithm~\ref{algalg2}
is a $\pi^{\ast}$-reversible Markov chain.
\end{prop}

%
\begin{example}[(Randomized MCMC \cite{nichollscoupled})]
In \cite{nichollscoupled}, the authors use the terminology \emph
{Randomized MCMC} (r-MCMC) for a $\pi^{\ast}$-reversible
Metropolis--Hastings chain $\sequence{\Y{}}$ generated using a set of
auxiliary variables\break  $\sequence{\Aux{}}$ with a particular expression of
the acceptance probability. Although only one of these auxiliary
variables is sampled at each time step, one may actually cast this
approach into the framework of Algorithm~\ref{algalg2} by creating
artificially another auxiliary variable according to the deterministic kernel
\[
\label{eqrmcmcT} T(y, u, \hat{y}; \mathrm{d}\hat{u}) = \delta _{f(u)}(
\mathrm{d}\hat{u}),
\]
where $f$ is any continuously differentiable involution on $\mathsf{U}$.
Even though $T$ is not dominated, it is possible to verify (\ref
{eqcondradon}) using that $f$ is an involution. We prove in
Appendix~\ref{apprMCMC} that the r-MCMC algorithm is a special case of
Algorithm~\ref{algalg2} with this particular choice of $T$ and with
the general form of the acceptance probability described in Remark~\ref
{remgeneralradon}.
\end{example}

%
\begin{example}[(Generalized Multiple-try Metropolis \cite
{pandolfigeneralization})]
The \emph{Generalized Multiple-try Metropolis} (GMTM) \emph{algorithm}
\cite{pandolfigeneralization} is an extension of the \emph
{Multiple-try Metropolis--Hastings algorithm} proposed in \cite
{liumultiple}. Given $\Y[k]{}=y$, one draws $n$ {i.i.d.}
possible moves
$V_1, \ldots, V_n$ according to $\check{R}(y, \cdot)$. After
this, a random index $J$ taking the value $j \in\{1,\ldots,n\}$ with
probability proportional to $\omega(y, V_{j})$ is generated,
whereupon a candidate is constructed as $\tY= V_J$. The candidate
is then accepted with some probability that is computed using $n$
additional random variables $\tAuxiv_1, \ldots, \tAuxiv_n$, where
$\tAuxiv_1,\ldots, \tAuxiv_{n-1}$ are i.i.d. draws from $\check
{R}(\hat{y},
\cdot)$, and $\tAuxiv_n$ is set deterministically to $\tAuxiv_n=y$
(see Appendix~\ref{appMTM} for more details concerning the acceptance
probability). In Appendix~\ref{appMTM}, Proposition~\ref{lemMTM}, it is shown
that the GMTM algorithm is in fact a special case\vspace*{2pt} of Algorithm~\ref
{algalg2} with $U= (V_1, \ldots, V_{J - 1}, V
_{J +
1}, \ldots, V_n)$ and $\tAuxi= (\tAuxiv_1, \ldots, \tAuxiv
_{n - 1})$.
\end{example}

When the function $k \dvtx(y, \hat{y}) \mapsto\int R(y, \mathrm{d}u)
s(y, u; \hat{y})$ is known explicitly, one may obtain another
$\pi^{\ast}
$-reversible Markov chain by means of the classical
Metropolis--Hastings ratio, that is, we use again Algorithm~\ref
{algalg2} but replace the acceptance probability $\alpha(y, u, \hat{y}, \hat{u})$ by
%
\begin{equation}
\label{eqMHratio} \talpha(y,\hat{y}):=1\wedge\frac{\pi^{\ast}
(\hat{y})k(\hat{y}, y)}{\pi^{\ast}(y
)k(y, \hat{y})}.
\end{equation}
The following proposition, which generalizes a similar result obtained
in \cite{nichollscoupled}, Section~2.3, for the r-MCMC algorithm,
shows, when combined with \cite{tierneynote}, Theorem~4, that the
asymptotic variance of the classical Metropolis--Hastings estimator is
smaller than that of the estimator based on Algorithm~\ref{algalg2}.
%
\begin{prop} \label{propMHAlg2}
The Metropolis--Hastings kernel associated with the acceptance
probability (\ref{eqMHratio}) is larger, in the sense of
Definition~\ref{defiPeskunordering}, than the transition kernel
associated with Algorithm~\ref{algalg2}.
\end{prop}

\begin{pf}
Set
\[
\mu( \mathrm{d}u\times\mathrm{d}\hat{u}):=\frac{R(y, \mathrm{d}u)
s(y, u; \hat{y}) T(y, u, \hat{y}; \mathrm{d}\hat{u})}{k(y, \hat{y})}
\]
and note that $\mu$ is a probability measure. Hence, as the mapping
$\mathbb{R} \ni v \mapsto1 \wedge v$ is concave, Jensen's inequality
implies that
\begin{eqnarray*}
&& \frac{ \iint R(y,\mathrm{d}u) s(y,u;\hat{y})
T(y,u,\hat{y};\mathrm{d}\hat{u}) \alpha(y,u,\hat{y},\hat{u}) }{ k(y,\hat{y})\talpha(y,\hat{y})}
\\
&&\qquad =\frac{\iint\mu(\mathrm{d}u\times\mathrm{d}\hat{u}) \alpha(y, u, \hat{y}, \hat{u})}{\talpha(y, \hat{y})}
\\
&&\qquad \leq  \biggl( 1 \wedge\iint\mu(\mathrm{d}u\times\mathrm {d}\hat{u})
\frac{\pi(\hat{y}, \hat{u}) s(\hat{y}, \hat{u}; y) t(\hat
{y}, \hat{u}, y; u
)}{\pi(y, u) s(y, u;\hat{y}) t(y, u, \hat{y}; \hat{u})} \biggr) \Big/ \talpha(y, \hat{y})
\\
&&\qquad = 1 \wedge\frac{\pi^{\ast}(\hat{y})k(\hat{y}, y)}{\pi
^{\ast}
(y)k(y, \hat{y})} \Big/ \talpha(y, \hat{y}) = 1
\end{eqnarray*}
(a similar technique was used in the proof of \cite
{andrieuconvergence}, Lemma~1). The previous computation shows that
the off-diagonal transition density function of the
Metropolis--Hastings Markov chain associated with the acceptance
probability (\ref{eqMHratio}) is larger than that of the chain in
Algorithm~\ref{algalg2}. This completes the proof.
\end{pf}
However, in practice a closed-form expression of $k$ is rarely
available, which prevents the classical Metropolis--Hastings algorithm
from being implemented. Thus, if the transition density $r$ is known
explicitly and can be sampled we have to choose between Algorithms~\ref{algalg1}~and~\ref{algalg2} for approximating $\pi^{\ast
}$. The
classical tools (such as the ordering in Definition~\ref
{defiPeskunordering}) for comparing \sequence{\Y{1}} and \sequence
{\Y
{2}} cannot be applied here, since \sequence{\Y{1}} is not even a
Markov chain. Nevertheless, Theorem~\ref{teomainResult} allows these
two algorithms to be compared theoretically by embedding
\sequence{\Y{1}} and \sequence{\Y{2}} into inhomogeneous $\pi
$-reversible Markov chains. The construction, which will be carried
through in full detail below, leads to the following result.

%
\begin{teo} \label{teocompAlg1Alg2}
Let \sequence{\Y{1}} and \sequence{\Y{2}} be sequences of random
variables generated by Algorithms \ref{algalg1}~and~\ref{algalg2}, respectively,
where $(\Y[0]{1},\break  \Aux[0]{1}) \sim\pi$ and
$\Y[0]{2} \sim\pi^{\ast}$. Then for all $h \in\Ltwo[\pi^{\ast
}]$ satisfying
%
\begin{equation}
\label{eqcondh} \sum_{k = 1}^\infty\bigl| \covardu{h
\bigl(\Y[0]{i} \bigr)} {h \bigl(\Y[k]{i} \bigr)} \bigr| < \infty\qquad \bigl(i \in\{1,2
\} \bigr)
\end{equation}
it holds that
\[
\lim_{n \to\infty} \frac{1}{n} \var{\sum
_{k = 0}^{n - 1} h \bigl(\Y[k]{2} \bigr)} \leq\lim
_{n \to\infty} \frac{1}{n} \var{\sum_{k = 0}^{n - 1}
h \bigl(\Y[k]{1} \bigr)}.
\]
\end{teo}

We preface the proof of Theorem~\ref{teocompAlg1Alg2} by the following
lemma, which may serve as a basis for the comparison of \emph
{homogeneous} Markov chains evolving according to $P_i Q_i$ (or $ Q_i
P_i$), $i \in\{0, 1\}$, where $P_i$ and $ Q_i$, $i \in\{0, 1\}$, are
kernels satisfying~\hyperref[assPeskunorder]{(A1)} on some product space.

%
\begin{lem} \label{lemhomogeneouscomp}
Let $P_i$ and $ Q_i$, $i \in\{0, 1\}$, be kernels satisfying \textup{\hyperref
[assPeskunorder]{(A1)}} on $(\mathsf{X}, \mathcal{X})$, with
$\mathsf{X} = \mathsf{Y}
\times\mathsf{U}$ and $\mathcal{X}= \mathcal{Y}\varotimes\mathcal
{U}$. In
addition, assume that for all $(y, u) \in\mathsf{X}$,
%
\begin{equation}
\label{eqassunitmass} P_i \bigl(y, u; \{ y \} \times\mathsf{U} \bigr) = 1
\qquad \bigl(i \in\{0, 1\} \bigr).
\end{equation}
Then for all $f \in\Ltwo[\pi]$ depending on only the first argument
[i.e., $f(y, u) = h(y)$ for some $h$] and such that
%
\begin{equation}
\label{eqassbddnessproduct} \sum_{n = 1}^\infty\bigl| \bigl
\langle f, (P_i Q_i)^n f \bigr\rangle\bigr| <
\infty\qquad \bigl(i \in\{0, 1\} \bigr)
\end{equation}
it holds that
\[
v(f,P_1 Q_1) = v(f, Q_1 P_1)
\leq v(f, P_0 Q_0) = v(f, Q_0
P_0).
\]
\end{lem}

%
\begin{rem} \label{remcounterexlemma}
Assumption (\ref{eqassunitmass}) is essential in Lemma~\ref
{lemhomogeneouscomp}. Indeed, let $\mathsf{X} = \{-1, 1\}$ and $\pi(\{
1 \}) = \pi(\{ -1 \}) = 1/2$, and define the kernels $P_0(x, \mathrm{d}x)
= \delta_x(\mathrm{d}x')$, $ Q_0(x, \mathrm{d}x') = \varepsilon\pi
(\mathrm{d}x') +
(1 - \varepsilon) \delta_{-x}(\mathrm{d}x')$ for some $\varepsilon
\in(0,
1)$, $P_1(x, \mathrm{d}x') = \pi(\mathrm{d}x')$, and $ Q_1 = Q_0$.
Then the
kernels $P_i$ and $ Q_i$, $i \in\{0, 1\}$, satisfy
\hyperref[assPeskunorder]{(A1)}, and consequently Theorem~\ref
{teomainResult} applies to the inhomogeneous chains evolving
alternatingly according to the same. However, the similar result does
not hold true for chains evolving according to the product kernels $P_i
Q_i$ and $ Q_i P_i$, $i \in\{0, 1\}$, as
\[
v(f, P_0 Q_0) = v(f, Q_0
P_0) = \frac{\varepsilon}{2 - \varepsilon} < 1 = v(f,P_1
Q_1) = v(f, Q_1 P_1 ),
\]
with $f$ being the identity mapping on $\mathsf{X}$.
\end{rem}

\begin{pf*}{Proof of Lemma~\ref{lemhomogeneouscomp}}
Define Markov chains $\sequence{\X{i}}$, $i \in\{0, 1\}$, evolving as
\[
\label{eqXi} \cdots\stackrel{ Q_i} {\longrightarrow} \X[2k]{i} =
\pmatrix{\Y[k]{i}
\vspace*{3pt}\cr
\Aux[k]{i}}
\stackrel{P_i} {
\longrightarrow} \X[2k+1]{i}= %
\pmatrix{\cY[k]{i}
\vspace*{3pt}\cr
\cAux[k]{i} }
\stackrel{ Q_i} {
\longrightarrow} \X[2k+2]{i}= %
\pmatrix{\Y[k+1]{i}
\vspace*{3pt}\cr
\Aux[k+1]{i}}
\stackrel{P_i} {
\longrightarrow} \cdots
\]
with $\X[0]{i} \sim\pi$. By construction,
%
\begin{eqnarray}
\label{eqimpliedbddness}
\qquad && \sum_{k = 1}^{\infty} \bigl( \bigl|
\covardu{f \bigl(\X[0]{i} \bigr)} {f \bigl(\X[k]{i} \bigr)}\bigr|+\bigl|\covardu {f \bigl(
\X[1]{i} \bigr)} {f \bigl(\X[k+1]{i} \bigr)}\bigr| \bigr)
\nonumber\\[-8pt]\\[-8pt]
&&\qquad = \pi f^2 - \pi^2 f + 4 \sum
_{k = 1}^\infty\bigl|\covardu{h \bigl(\Y[0]{i} \bigr)} {h \bigl(
\Y[k]{i} \bigr)}\bigr| < \infty\qquad \bigl(i \in\{0, 1\} \bigr), \nonumber
\end{eqnarray}
where finiteness follows from the assumption (\ref
{eqassbddnessproduct}). Moreover, for all $n \in\mathbb{N}^{\ast}$
and $i \in\{
0, 1\}$,
\[
\var{\sum_{k = 0}^{n - 1} h \bigl(\Y[k]{i}
\bigr)} = \var{\sum_{k = 0}^{n - 1} h \bigl(
\cY[k]{i} \bigr)} = \frac{1}{4} \var{\sum_{k = 0}^{2n - 1}
f \bigl(\X[k]{i} \bigr)},
\]
which implies, by (\ref{eqimpliedbddness}),
\[
v(f, P_i Q_i) = v(f, Q_i
P_i) = \frac{1}{2} \lim_{n \to\infty}
\frac
{1}{n} \var{\sum_{k = 0}^n f
\bigl(\X[k]{i} \bigr)} \qquad \bigl(i \in\{0, 1\} \bigr).
\]
Finally, by (\ref{eqimpliedbddness}) we may now apply Theorem~\ref
{teomainResult} to the chains $\{ \X[k]{i}; k \!\in\!\mathbb{N}\}$, $i
\in\{
0, 1\}$, which establishes immediately the statement of the lemma.
\end{pf*}

\begin{pf*}{Proof of Theorem~\ref{teocompAlg1Alg2}}
We introduce the kernels:
\begin{itemize}
\item$P_1(y, u; \mathrm{d}y' \times\mathrm{d}u') = \delta_{(y,
u)}(\mathrm{d}y'
\times\mathrm{d}u')$,
\item$P_2(y, u; \mathrm{d}y' \times\mathrm{d}u') = \delta
_y(\mathrm{d}y') R(y,
\mathrm{d}u')$,
\item$ Q_1= Q_2$ being-defined implicitly as the transition kernel
associated with the \emph{freeze algorithm} (Algorithm~\ref{algalg1}).
\end{itemize}
It can be checked readily that the two sequences $\sequence{\Y{1}}$ and
$\sequence{\Y{2}}$ generated by Algorithms~\ref{algalg1}~and~\ref{algalg2},
respectively, have indeed the same
distributions as the marginal processes (with respect to the first
component) of homogeneous chains evolving according to the products
$P_1 Q_1$ and $P_2 Q_2$, respectively. In addition, all kernels $P_i$
and $ Q_i$, $i \in\{1, 2\}$, are $\pi$-reversible, as:
\begin{itemize}
\item$P_1$ is reversible with respect to any probability measure (in
particular, it is \mbox{$\pi$-}reversible),
\item$P_2$ is $\pi$-reversible as a Gibbs-sampler sub-step
transition kernel,
\item$ Q_1= Q_2$ is $\pi$-reversible as a classical
Metropolis--Hastings transition kernel.
\end{itemize}
Since $P_1$ has no off-diagonal component, it holds that $P_2 \succeq
P_1$; moreover, trivially, $ Q_2 = Q_1 \succeq Q_1$. Thus, we may
complete the proof by applying Lemma~\ref{lemhomogeneouscomp} to the
function $f(y, u) = h(y)$, for which the condition (\ref
{eqassbddnessproduct}) is satisfied [by (\ref{eqcondh})].
\end{pf*}


\subsection*{Case \textup{II}: Simulating $R$-transitions is infeasible}

\emph{Pseudo-marginal algorithms} (see \cite{andrieupseudo} and \cite
{andrieuconvergence}) are implemented using a Markov kernel $\check
{R}$ on
$\mathsf{Y}\times\mathcal{U}$ and a family $\{w_u; u\in\mathsf
{U}\}$
of real-valued nonnegative functions on $\mathsf{Y}$ such that $\int
\check{R}(y,\mathrm{d}u) w_{u}(y)=1$ for all $y\in\mathsf{Y}$. We denote by
$\check{r}$ the transition density of the kernel $\check{R}$ with
respect to
some dominating measure. Note that $R(y, \mathrm{d}u):=\check
{R}(y,\mathrm{d}
u) w_{u}(y)$ is a Markov transition kernel as well. The
problem at hand is to sample the target distribution
\[
\pi(\mathrm{d}y\times\mathrm{d}u):=\pi^{\ast}(\mathrm{d}y) R(y,
\mathrm{d}u) = \pi^{\ast}(\mathrm{d}y) \check{R}(y, \mathrm{d}u)
w_{u}(y)
\]
under the assumption that:
\begin{itemize}
\item for all $(y, u) \in\mathsf{Y}\times\mathsf{U}$, $\pi^{\ast
}(y) \check
{r}(y, u) w_{u}(y)$ is known up to a normalizing constant,
\item for all $y\in\mathsf{Y}$, $\check{R}(y, \cdot)$ can be
sampled from.\vadjust{\goodbreak}
\end{itemize}
The particular case where $w_{u}(y) = 1$ for all $(y, u) \in
\mathsf{Y}\times\mathsf{U}$ was discussed in the previous section,
and we now
turn to the case $w_{u}(y)\neq1$ (i.e., sampling directly from
$R$ is infeasible). The solution provided by pseudo-marginal algorithms
consists in replacing, in Algorithm~\ref{algalg2}, the operation~(i)
by the sampling $U\sim\check{R}(y, \cdot)$, and the computing the
acceptance probability $\alpha$ [as defined in (\ref
{eqacceptMetropolis})] via the formula
\[
\alpha(y, u, \hat{y}, \hat{u}):=1\wedge\frac{\pi^{\ast}(\hat
{y}) \check
{r}(\hat{y}, \hat{u}) w_{\hat{u}}(\hat{y}) s(\hat{y}, \hat{u}; y
) t(\hat{y}, \hat{u}, y;
u)}{\pi^{\ast}(y) \check{r}(y, u) w_{u}(y) s(y, u; \hat{y})
t(y, u, \hat{y}; \hat{u})}.
\]
The output of this algorithm, which will be referred to as the \emph
{noisy algorithm} in the following, is typically not---on the contrary
to Algorithm~\ref{algalg2}---\break $\pi^{\ast}$-reversible due to the
replacement of $R$ by $\check{R}$. This justifies the denomination. However,
when $w$ is close to unity the noisy algorithm is close to
Algorithm~\ref{algalg2}, which is, according to Theorem~\ref
{teocompAlg1Alg2}, more efficient than Algorithm~\ref{algalg1} in
terms of asymptotic variance.

%
\begin{example}[(Monte Carlo Within Metropolis)]
The Monte Carlo Within Metropolis algorithm (MCWM; see \cite
{andrieupseudo}) resembles closely the GIMH algorithm (see Example \ref
{exgimh}), however, with the important difference that the importance
sampling estimates $\pi_N^*(Y_k)$ [given by (\ref{eqGIMHMCest})]
are \emph{not} stored and propagated through the algorithm along with
the $Y_k$-values. Instead each estimate of the marginal density is
recomputed using a ``fresh'' MC sample before the calculation of the
acceptance probability. Thus, the MCWM algorithm can be cast into the
framework of the noisy algorithm with $T = \check{R}$ and with the auxiliary
variables $U$ and $\tAuxi$ playing the roles of $N$-dimensional
Monte Carlo samples.
\end{example}

Considering this, we now propose a novel algorithm which will be
referred to as the \textit{random refreshment} \emph{algorithm} and
which is a hybrid between Algorithm~\ref{algalg2} and the noisy
algorithm. This novel algorithm, which is described in Algorithm~\ref
{algalgRefresh} below, targets \emph{exactly} $\pi^{\ast}$ and turns out
to be more efficient than Algorithm~\ref{algalg1}.

\begin{algorithm}[h]
\begin{algorithmic}[7]
\caption{The \textit{random refreshment} algorithm} \label{algalgRefresh}
\Require$(\Y[k]{3}, \Aux[k]{3}) = (y, u)$:
\begin{longlist}[(iii)]
\item[(i)]
\begin{itemize}[(i.1)]
\item[(i.1)] draw $U' \sim\check{R}(y, \cdot) \leadsto u'$,
\item[\hspace*{34pt}(i.2)] set
%
\begin{equation}
\label{eqdefrho} \check{U}\gets%
\cases{
u', &\quad with probability $\varrho \bigl(y, u, u'\bigr):=1 \wedge\dfrac{w_{u'}(y)}{w_{u}(y)}$,
\vspace*{3pt}\cr
u, &\quad otherwise,}\hspace*{16pt}\leadsto\check{u},
\end{equation}
\end{itemize}
\item[(ii)] draw $\tY\sim S(y, \check{u}; \cdot) \leadsto\hat{y}$,
\item[(iii)] draw $\tAuxi\sim T(y, \check{u}, \hat{y}; \cdot) \leadsto
\hat{u}$,
\item[(iv)] set
$  (\Y[k+1]{3}, \Aux[k+1]{3}) \gets
\cases{
(\hat{y}, \hat{u}), &\quad with probability $\alpha(y, \check {u}, \hat{y}, \hat{u})$,
\vspace*{3pt}\cr
(y, \check{u}), &\quad otherwise.}
$
\end{longlist}
\end{algorithmic}
\end{algorithm}

In step~(i) in Algorithm~\ref{algalgRefresh}, the auxiliary variable
$\check{U}$ can be either ``refreshed,'' that is, replaced by a new
candidate $U'$, or kept at the previous state $\Aux[k]{3}$
according to an acceptance probability that turns out to be a standard
Metropolis--Hastings acceptance probability (which will be seen in the
proof of Theorem~\ref{teocompAlg1Alg4} below). Interestingly,\vspace*{1pt} this
allows the desired distribution $\pi$ as the target distribution of
$\{(\Y[k]{3}, \Aux[k]{3}); k \in\mathbb{N}\}$. In comparison, the noisy
algorithm described above differs only from Algorithm~\ref
{algalgRefresh} by step~(i), in that the new candidate is always
accepted in the noisy algorithm. This ``systematic refreshment'' makes
actually the noisy algorithm imprecise in the sense that $\pi$ is no
longer the target distribution except when $w_{u}(y)=1$ for all
$(y, u) \in\mathsf{Y}\times\mathsf{U}$, in which case $\varrho(y,u,u
')$ in (\ref{eqdefrho}) becomes identically equal to unity and
Algorithm~\ref{algalgRefresh} translates into Algorithm~\ref
{algalg2}. Compared to Algorithm~\ref{algalg1}, step~(i) allows the
second component to be refreshed randomly according to the probability
$\varrho(y,u,\check{u})$ whereas this component remains unchanged in
Algorithm~\ref{algalg1}. Thus, in conformity with Algorithm~\ref
{algalg2}, it is likely that Algorithm~\ref{algalgRefresh} has better
mixing properties than Algorithm~\ref{algalg1}. That this is indeed
the case may be established by reapplying the embedding technique
developed in the previous part. Before formalizing this properly, we
propose an example showing a typical situation where a Random
Refreshment algorithm may be used.

%
\begin{example}[(Random refreshment GIMH-ABC)]
\label{exrrABC}
In \cite{leeandrieudoucet2012} (contributing to the
discussion of \cite{fearnhead2012constructing}), the authors propose a
novel algorithm, \emph{rejuvenating GIMH-ABC}~\cite
{leeandrieudoucet2012}, Algorithm~1, preventing the original \emph
{GIMH-ABC}~\cite{fearnhead2012constructing}, Algorithm~2 (termed \emph
{MCMC-ABC} in the paper in question), from falling into possible
trapping states. The GIMH-ABC is an instance of Algorithm~\ref
{algalg1} targeting $\pi(\mathrm{d}y\times\mathrm{d}u\mid\obs):=
\pi^{\ast}(\mathrm{d}y\mid\obs) \check{R}(y, \mathrm{d}u)
w_{u}(y,\obs
)$, where, in the ABC context:
\begin{itemize}
\item$\pi^{\ast}(\mathrm{d}y\mid\obs)$ is the desired posterior
of a
parameter $y$ given some observed data summary statistics $\obs$,
\item$\check{R}(y, \cdot)$ is the likelihood of the data (from which
sampling is assumed to be feasible),
\item$w_{u}(y, \obs):=K[(s(u)-\obs)/h] / \int\check{R}(y,
\mathrm{d}u') K[(s(u') - \obs) / h ]$, where $K$ is a kernel
integrating to unity, providing the classical ABC discrepancy measure
between the observed data summary statistics $\obs$ and that evaluated
at the simulated data $u$.
\end{itemize}
Rejuvenating GIMH-ABC comprises an intermediate step in which the
simulated data $u$, generated under the current parameter $y$, are
refreshed systematically. However, since sampling from $R(y, \mathrm{d}u
):=\check{R}(y, \mathrm{d}u) w_{u}(y,\obs)$ is typically
infeasible, the auxiliary variables are refreshed through $\check{R}$
in the
spirit of Algorithm~\ref{algalg2}. Therefore, in accordance with
Algorithm~\ref{algalgRefresh}, a \emph{$\pi$-reversible} alternative
to rejuvenating GIMH-ABC is obtained by, instead of refreshing
systematically the data, performing refreshment with probability (\ref
{eqdefrho}). Note that the fact that the constant in the denominator
of $w_{u}(y,\obs)$ is typically not computable does not prevent
computation of (\ref{eqdefrho}), since this constant appears in
$w_{u}(y,\obs)$ as well as $w_{u'}(y,\obs)$. This provides a
\emph{random refreshment GIMH-ABC}, which can be compared
quantitatively, via the Theorem~\ref{teocompAlg1Alg4} below, to the
GIMH-ABC while at the same time avoiding the possible GIMH-ABC trapping
states mentioned in \cite{leeandrieudoucet2012}.
\end{example}

%
\begin{teo} \label{teocompAlg1Alg4}
Let \sequence{\Y{1}} and \sequence{\Y{3}} be the sequences of random
variables generated by Algorithms~\ref{algalg1}~and~\ref{algalgRefresh},
respectively, where $(\Y[0]{i},\break  \Aux[0]{i}) \sim
\pi
$, $i \in\{1, 3\}$. Then the following hold true:
\begin{longlist}[(ii)]
\item[(i)] The output of Algorithm~\ref{algalgRefresh} is $\pi$-reversible.
\item[(ii)] For all $h \in\Ltwo[\pi^{\ast}]$ satisfying
\[
\label{eqcondhalg1alg4} \sum_{k = 1}^{\infty} \bigl| \covardu{h
\bigl(\Y[0]{i} \bigr)} {h \bigl(\Y[k]{i} \bigr)} \bigr| < \infty\qquad \bigl(i \in\{1,3
\} \bigr)
\]
it holds that
\[
\lim_{n \to\infty} \frac{1}{n} \var{\sum
_{k = 0}^{n - 1} h \bigl(\Y[k]{3} \bigr)} \leq\lim
_{n \to\infty} \frac{1}{n} \var{\sum_{k = 0}^{n - 1}
h \bigl(\Y[k]{1} \bigr)}.
\]
\end{longlist}
\end{teo}

\begin{pf}
Let the kernels $P_1$ and $ Q_1$ be defined as in the proof of
Theorem~\ref{teocompAlg1Alg2} and introduce furthermore:
\begin{itemize}
\item$P_3$ defined implicitly by the transition $(\Y[k]{3}, \Aux
[k]{3}) \rightarrow(\Y[k]{3}, \check{U})$ according to step~(i) in
Algorithm~\ref{algalgRefresh} (note that the first component is held
fixed throughout the transition),
\item$ Q_3 = Q_1$.
\end{itemize}
In conformity with the proof of Theorem~\ref{teocompAlg1Alg2}, it can
be checked readily that the two sequences $\sequence{\Y{1}}$ and
$\sequence{\Y{3}}$ generated by Algorithms~\ref{algalg1}~and~\ref{algalgRefresh}, respectively, have indeed the same
distributions as the marginal processes (with respect to the first
component) of homogeneous chains evolving according to the products
$P_1 Q_1$ and $P_3 Q_3$, respectively. The $\pi$-reversibility of the
kernels $P_1$ and $ Q_1 = Q_3$ was established in the proof of
Theorem~\ref{teocompAlg1Alg2}. To verify $\pi$-reversibility of~$P_3$ as well, note that $P_3$ is a Metropolis--Hastings kernel
associated with the target distribution $\pi$, whose acceptance
probability includes a Radon--Nikodym derivative of the type given in
Proposition~\ref{propderiveeRadon}; it is therefore $\pi
$-reversible. Indeed, note that $P_3$ updates only the second component
according to $\check{R}(y, \mathrm{d}u')$ with the acceptance probability
$\varrho(y, u, u')$. Assuming first that $\check{R}$ is dominated and
denoting by $\check{r}$ its transition density, we have
\[
\varrho \bigl(y,u,u' \bigr) = 1 \wedge\frac{w_{u'}(y)}{w_{u}(y)} = 1 \wedge
\frac{\pi(y, u') \check{r}(y, u)}{\pi(y, u) \check{r}(y, u')},
\]
where $\pi(y, u) = \pi^{\ast}(y) \check{r}(y, u) w_{u}(y)$ in
the density of the target $\pi$. This shows that $\varrho(y, u,
u')$ is indeed the acceptance probability of a Metropolis--Hastings
Markov chain targeting $\pi$, with proposal kernel $\check{R}(y,
\mathrm{d}u') \delta_y(\mathrm{d}y')$; the $\pi$-reversibility of $P_3$ follows.
The proof can be adapted easily to the case where $\check{R}$ is not
dominated. As a consequence, the product $P_3 Q_3$ is also $\pi
$-reversible, which establishes the statement (i) of the theorem.
Finally, since $P_1$ has zero mass on the off-diagonal, it holds that
$P_3 \succeq P_1$ and, clearly, $ Q_3 = Q_1\succeq Q_1$. The proof of
(ii) is now concluded by applying Lemma~\ref{lemhomogeneouscomp}
along the lines of the proof of Theorem~\ref{teocompAlg1Alg2}.
\end{pf}

\section{Proof of Theorem~\texorpdfstring{\protect\ref{teomainResult}}{4}} \label
{secproofmain}We preface the proof of Theorem~\ref{teomainResult} with
some preliminary lemmas.
%
\begin{lem}
\label{lemlem1}
Assume that $P_{1},P_{2},\ldots,P_{n}$ are $\pi$-reversible Markov
transition kernels. Then, for all $(f,g)\in\Ltwo[\pi] \times\Ltwo
[\pi]$,
\[
\langle f,  P_{1}P_{2}\cdots P_{n}g \rangle=
\langle P_{n}\cdots P_{2} P_{1}f,  g \rangle.
\]
\end{lem}
\begin{pf}
As each $ P_{\ell}$ is $\pi$-reversible, it holds that $ \langle
P_{\ell}f,  g \rangle= \langle f,  P_{\ell}g
\rangle$ for all $(f, g) \in\Ltwo
[\pi]
\times\Ltwo[\pi]$ and $\ell\in\{1,\ldots,n\}$. Applying repeatedly
this relation yields
\begin{eqnarray*}
\langle f,  P_{1} P_{2}\cdots P_{n}g \rangle&=&
\langle P_{1}f,  P_{2}\cdots P_{n}g \rangle
\\
&=& \langle P_{2} P_{1}f,  P_{3}\cdots
P_{n}g \rangle= \cdots = \langle P_{n}\cdots P_{2}
P_{1}f,  g \rangle.
\end{eqnarray*}\upqed
\end{pf}


%
\begin{lem} \label{lemlem2}
Let $ P$ and $Q$ be Markov transition kernels on $(\mathsf{X},
\mathcal{X})$ such
that $\pi P=\pi Q=\pi$ and let $\sequence{\X{}}$ be a Markov
chain evolving as
\[
\X[0]{} \stackrel{ P} {\longrightarrow} \X[1]{} \stackrel{Q} {\longrightarrow}
\X[2]{} \stackrel{ P} {\longrightarrow} \X[3]{} \stackrel{Q} {\longrightarrow}
\cdots
\]
with initial distribution $\X[0]{}\sim\pi$. Then, for all $f\in
\Ltwo
[\pi]$ such that
%
\begin{equation}
\label{eqeq2assLem2} \sum_{k = 1}^{\infty} \bigl( \bigl|
\covardu{f \bigl(\X[0]{} \bigr)} {f \bigl(\X[k]{} \bigr)}\bigr|+\bigl|\covardu {f \bigl(\X[1]{}
\bigr)} {f \bigl(\X[k+1]{} \bigr)}\bigr| \bigr) <\infty,
\end{equation}
the limit, as $n$ tends to infinity, of ${n}^{-1} \varTxt{\sum_{k =
0}^{n - 1} f(\X[k]{})}$ exists, and
%
\begin{eqnarray}
\label{eqasvaraltexpression}
&& \lim_{n\to\infty}\frac{1}{n}\var {\sum
_{k = 0}^{n - 1} f \bigl(\X[k]{} \bigr)}\nonumber
\\
&&\qquad =  \pi
f^2 - \pi^2 f
\\
&&\quad\qquad{} +\sum_{k = 1}^{\infty} \covardu{f \bigl(\X[0]{}
\bigr)} {f \bigl(\X[k]{} \bigr)} +\sum_{k = 1}^{\infty}
\covardu{f \bigl(\X[1]{} \bigr)} {f \bigl(\X[k+1]{} \bigr)}. \nonumber
\end{eqnarray}
\end{lem}

\begin{pf}
As covariances are symmetric,
\[
\frac{1}{n} \var{\sum_{k = 0}^{n - 1} f
\bigl(\X[k]{} \bigr)}=\pi f^{2} - \pi^{2}f + 2
n^{-1} \sum_{0 \leq i < j \leq n - 1}\covardu{f \bigl(\X[i]{}
\bigr)} {f \bigl(\X[j]{} \bigr)}.
\]
We now consider the limit, as $n$ tends to infinity, of the last term
on the right-hand side. Let $\mathcal{E}$ and $\mathcal{O}$ denote
the two
complementary subsets of $\mathbb{N}$ consisting of the even and odd
numbers, respectively. For all $(i, j) \in\mathbb{N}^2$ such that $i
< j$,
we have
\[
\covardu{f \bigl(\X[i]{} \bigr)} {f \bigl(\X[j]{} \bigr)} = %
\cases{
\covardu{f \bigl(\X[0]{} \bigr)} {f \bigl(\X[j-i]{} \bigr)}, &\quad if $i \in\mathcal{E}$,
\vspace*{3pt}\cr
\covardu{f \bigl(\X[1]{} \bigr)} {f \bigl(\X[j-i+1]{} \bigr)}, &\quad if $i \in \mathcal{O}$.}
\]
This implies that
\begin{eqnarray*}
&& n^{-1} \mathop{\sum_{0\leq i<j\leq n-1}}_{i\in\mathcal{E}} \covardu{f \bigl(\X[i]{}
\bigr)} {f \bigl(\X[j]{} \bigr)}
\\
&&\qquad =
\sum_{k = 1}^{n - 1} n^{-1} \biggl(
\biggl\lfloor\frac{n - 1 - k}{2} \biggr\rfloor+ 1 \biggr) \covardu {f \bigl(\X[0]{}
\bigr)} {f \bigl(\X[k]{} \bigr)}
\end{eqnarray*}
and
\begin{eqnarray*}
&& n^{-1} \mathop{\sum_{0\leq i < j \leq n - 1}}_{i \in\mathcal{O}} \covardu{f \bigl(\X[i]{}
\bigr)} {f \bigl(\X[j]{} \bigr)}
\\
&&\qquad =\sum_{k = 1}^{n - 2} n^{-1} \biggl(
\biggl\lfloor\frac{n - 2 - k}{2} \biggr\rfloor+ 1 \biggr) \covardu {f \bigl(\X[1]{}
\bigr)} {f \bigl(\X[k+1]{} \bigr)}.
\end{eqnarray*}
Under\vspace*{1pt} (\ref{eqeq2assLem2}), the dominated convergence theorem
applies, which provides that the limit, as $n$ goes to infinity, of
$n^{-1} \varTxt{\sum_{k = 0}^{n - 1} f(\X[k]{})}$ exists and is equal
to (\ref{eqasvaraltexpression}).
\end{pf}


%
\begin{lem} \label{lemlem4}
Let $P_i$ and $ Q_i$, $i\in\{0,1\}$, be $\pi$-reversible Markov
kernels on $(\mathsf{X},\mathcal{X})$ such that $P_0\pgeq P_1$ and $
Q_0\pgeq Q_1$.
For all $n\in\mathbb{N}$ and $i \in\{0, 1\}$, denote by $\R[n]{i}$ the
Markov kernel $\R[n]{i}:=P_i \mathbh{1}_{\mathcal{E}}(n) + Q_i
\mathbh{1}_{\mathcal{O}}(n)$. In addition, let $f\in\Ltwo[\pi]$ be such that
for $i
\in\{0, 1\}$,
%
\begin{equation}
\label{eqeq4assLem4} \sum_{k = 1}^{\infty} \bigl\vert \bigl
\langle f,  \R[0]{i} \cdots\R[k-1]{i}f \bigr\rangle \bigr\vert< \infty.
\end{equation}
Then for all $\lambda\in(0,1)$,
\begin{eqnarray*}
&& \sum_{k = 1}^{\infty} \lambda^k
\bigl( \bigl\langle f,  \R[0]{1} \cdots\R[k-1]{1}f \bigr\rangle + \bigl\langle f,
\R [1]{1} \cdots\R[k]{1} f \bigr\rangle \bigr)
\\
&&\qquad \leq\sum_{k=1}^{\infty} \lambda^k
\bigl( \bigl\langle f,  \R[0]{0} \cdots\R[k-1]{0}f \bigr\rangle + \bigl\langle f,
\R [1]{0} \cdots \R[k]{0}f \bigr\rangle \bigr).
\end{eqnarray*}
\end{lem}

\begin{pf}
For all $n\in\mathbb{N}$ and all $\alpha\in(0,1)$, define $\R
[n]{\alpha}:=(1-\alpha)\R[n]{0}+\alpha\R[n]{1}$. In addition,
set, for $\lambda\in(0,1)$, $\kH[\lambda]{}(\alpha):=
\kH
[\lambda]{\mathcal{E}}(\alpha)+\kH[\lambda]{\mathcal{O}}(\alpha
)$, where
\begin{eqnarray*}
\kH[\lambda]{\mathcal{E}}(\alpha) &:=&\sum
_{k = 1}^{\infty} \lambda^k \bigl\langle f,
\R[0]{ \alpha} \cdots \R[k-1]{\alpha} f \bigr\rangle,
\\
\kH[\lambda]{\mathcal{O}}(\alpha) &:=& \sum_{k = 1}^{\infty}
\lambda^k \bigl\langle f,  \R[1]{\alpha} \cdots\R[k]{\alpha }f \bigr
\rangle.
\end{eqnarray*}
Now, fix a distinguished $\lambda\in(0,1)$; we want show that
for all $\alpha\in[0,1]$,
%
\begin{equation}
\label{eqnegder} \frac{\mathrm{d}\kH[\lambda]{}}{\mathrm{d}\alpha
}(\alpha )\leq0.
\end{equation}
Thus, we start with differentiating $\kH[\lambda]{\mathcal{E}}$:
%
\begin{equation}
\label{eqderivH} \frac{\mathrm{d}\kH[\lambda]{\mathcal
{E}}}{\mathrm{d}\alpha
}(\alpha) = \frac{\mathrm{d}}{\mathrm{d}
\alpha} \sum
_{k = 1}^{\infty} \lambda^k \bigl\langle f,
\R[0]{ \alpha}\cdots \R[k-1]{\alpha}f \bigr\rangle.
\end{equation}
To interchange $\frac{\mathrm{d}}{\mathrm{d}\alpha}$ and $\sum_{k=1}^{\infty}$
in the
previous equation, we first note that
\begin{eqnarray*} \frac{\mathrm{d}}{\mathrm{d}\alpha} \bigl\langle f,  \R[0]{\alpha}
\cdots \R[k-1]{\alpha}f \bigr\rangle&=&  \sum
_{\ell= 0}^{k - 1} \frac{\partial}{\partial
\alpha_\ell}   \bigl\langle f, \R[0]{\alpha_0} \cdots\R [k - 1]{\alpha_{k - 1}}f \bigr
\rangle \bigg\vert_{(\alpha_0,\ldots,\alpha_{k -
1})=(\alpha,\ldots,\alpha)}
\\
&=&  \sum_{\ell= 0}^{k - 1} \bigl
\langle f,  \R[0 \nnearrow\ell-1]{\alpha} \bigl(\R[\ell]{1} - \R[\ell ]{0} \bigr)
\R[\ell+ 1 \nnearrow k - 1]{\alpha}f \bigr\rangle,
\end{eqnarray*}
where $\R[s\nnearrow t]{\alpha}:=\R[s]{\alpha} \R[s+1]{\alpha}
\cdots\R[t]{\alpha}$ for $s\leq t$ and $\R[s\nnearrow t]{\alpha
}:=
\operatorname{id}$ otherwise. By (\ref{eqmajoNormP}), $\|\R
[n]{\alpha
}\|\leq1$, which implies that\vadjust{\goodbreak} $\sup_{\alpha\in[0,1]}|\frac{\mathrm{d}
}{\mathrm{d}
\alpha} \langle f, \R[0]{\alpha}\cdots\R[k-1]{\alpha}f \rangle
|\leq2k
\pi(f^2)$. Thus, as $\sum_{k=1}^{\infty}\lambda^k k <\infty$ we may
interchange, in (\ref{eqderivH}), $\frac{\mathrm{d}}{\mathrm
{d}\alpha}$ and
$\sum_{k=1}^{\infty}$, yielding
\[
\frac{\mathrm{d}\kH[\lambda]{\mathcal{E}}}{\mathrm{d}\alpha
}(\alpha)=\sum_{k=1}^{\infty
}
\lambda^{k}\sum_{\ell=0}^{k-1} \bigl
\langle f,  \R[0 \nnearrow\ell-1]{\alpha} \bigl(\R[\ell ]{1}-\R [\ell]{0} \bigr)
\R[ \ell+1 \nnearrow k-1]{\alpha}f \bigr\rangle.
\]
Similarly, it can be established that
\[
\frac{\mathrm{d}\kH[\lambda]{\mathcal{O}}}{\mathrm{d}\alpha
}(\alpha) = \sum_{k=1}^{\infty
}
\lambda^{k}\sum_{\ell=1}^{k} \bigl
\langle f,  \R[1 \nnearrow\ell-1]{\alpha} \bigl(\R[\ell ]{1}-\R [\ell]{0} \bigr)
\R[\ell +1 \nnearrow k]{\alpha}f \bigr\rangle.
\]
We\vspace*{1pt} now apply Lemma~\ref{lemlem1} to the two previous sums. For this
purpose, we will use the following notation: $\R[s\ssearrow t]{\alpha
}:=\R[s]{\alpha} \R[s-1]{\alpha} \cdots\R[t]{\alpha}$ for
$s\geq
t$ and $\R[s\ssearrow t]{\alpha}:=\operatorname{id}$ otherwise.
Then
\begin{eqnarray*}
\frac{\mathrm{d}\kH[\lambda]{}}{\mathrm
{d}\alpha
}(\alpha)
&=& \sum_{k=1}^{\infty} \lambda^k
\Biggl\{\sum_{\ell=0}^{k-1} \bigl\langle\R[\ell-1
\ssearrow 0]{\alpha}f,  \bigl(\R[\ell]{1}-\R[\ell]{0} \bigr)\R [\ell+1 \nnearrow
k-1]{\alpha}f \bigr\rangle
\\[1pt]
&&\hspace*{33pt}{}+\sum_{\ell=1}^{k} \bigl\langle\R[\ell-1
\ssearrow1]{\alpha}f,  \bigl(\R[\ell]{1}-\R[\ell]{0} \bigr)\R [\ell+1 \nnearrow
k]{ \alpha}f \bigr\rangle \Biggr\}
\\[1pt]
&=& \sum_{\ell=0}^{\infty}\sum
_{m=0}^{\infty}\lambda^{\ell+m+1} \bigl\langle\R[
\ell-1 \ssearrow0]{\alpha}f,  \bigl(\R[\ell]{1}-\R[\ell]{0} \bigr)\R [\ell+1
\nnearrow \ell+m]{\alpha}f \bigr\rangle
\\[1pt]
&&{}{}+\sum_{\ell=1}^{\infty}\sum
_{m=1}^{\infty}\lambda^{\ell+m-1} \bigl\langle\R[
\ell-1 \ssearrow1]{\alpha}f,  \bigl(\R[\ell]{1}-\R[\ell]{0} \bigr)\R [\ell+1
\nnearrow \ell+m-1]{\alpha}f \bigr\rangle.
\end{eqnarray*}
Now, note that $\R[n]{\alpha}=\R[n']{\alpha}$ for all $(n, n') \in
\mathcal{O}
^2$ and $\R[m]{\alpha}=\R[m']{\alpha}$ for all $(m,m')^2 \in
\mathcal{E}^2$;
hence, separating, in the two previous sums, odd and even indices $\ell
$ provides
\begin{eqnarray*}
\frac{\mathrm{d}\kH[\lambda]{}}{\mathrm
{d}\alpha
}(\alpha)
&=& \sum_{\ell\in\mathcal{E}}\sum_{m=0}^{\infty}
\lambda^{\ell+m+1} \bigl\langle\R[1 \nnearrow\ell]{\alpha}f,  \bigl(\R[0]{1}-
\R[0]{0} \bigr)\R[1 \nnearrow m]{\alpha}f \bigr\rangle
\\[1pt]
&&{}+ \sum_{\ell\in\mathcal{E}\setminus\{0\}}\sum_{m=1}^{\infty}
\lambda^{\ell
+m-1} \bigl\langle\R[1 \nnearrow\ell-1]{\alpha}f,  \bigl(\R
[0]{1}- \R[0]{0} \bigr)\R[1 \nnearrow m-1]{\alpha}f \bigr\rangle
\\[1pt]
&&{}+ \sum_{\ell\in\mathcal{O}}\sum_{m=0}^{\infty}
\lambda^{\ell+m+1} \bigl\langle\R[0 \nnearrow\ell-1]{\alpha}f,  \bigl(
\R[1]{1}- \R[1]{0} \bigr)\R[0 \nnearrow m-1]{\alpha}f \bigr\rangle
\\[1pt]
&&{}+ \sum_{\ell\in\mathcal{O}}\sum_{m=1}^{\infty}
\lambda^{\ell+m-1} \bigl\langle\R[0 \nnearrow\ell-2]{\alpha}f,  \bigl(
\R[1]{1}- \R[1]{0} \bigr)\R[0 \nnearrow m-2]{\alpha}f \bigr\rangle.
\end{eqnarray*}
Finally, by combining the even and the odd sums,
\begin{eqnarray*}
\frac{\mathrm{d}\kH[\lambda]{}}{\mathrm
{d}\alpha
}(\alpha)
&=& \Biggl\langle\sum_{\ell=0}^{\infty}
\lambda^{\ell}\R[1 \nnearrow\ell]{\alpha}f,  \bigl(\R[0]{1}-\R[0]{0}
\bigr) \sum_{m=0}^{\infty} \lambda^m
\R[1 \nnearrow m]{\alpha}f \Biggr\rangle
\\
&&{}+ \Biggl\langle\sum_{\ell=0}^{\infty}
\lambda^\ell\R[0 \nnearrow \ell-1]{\alpha}f,  \bigl(\R[1]{1}-\R[1]{0}
\bigr) \sum_{m=0}^{\infty}\lambda^m\R
[0\nnearrow m-1]{\alpha}f \Biggr\rangle.
\end{eqnarray*}
Since $\R[n]{1}\pgeq[1] \R[n]{0}$, the operator $\R[n]{0}-\R
[n]{1}$ is
nonnegative on $\Ltwo[\pi]$ (by \cite{tierneynote}, Lemma 3), and
for all $f\in\Ltwo[\pi]$ it holds that $\langle f, (\R[n]{1}-\R
[n]{0})f \rangle\leq0$. This shows~(\ref{eqnegder}), which implies
that the function $\alpha\mapsto\kH[\lambda]{}(\alpha)$ is
nonincreasing on $(0,1)$. The proof is complete.
\end{pf}

\begin{pf*}{Proof of Theorem~\ref{teomainResult}}
According to Lemma~\ref{lemlem2}, for all functions $f\in\Ltwo[\pi
]$ and $i\in\{0,1\}$,
%
\begin{eqnarray}\label{eqth4proof1}
v^{(i)}(f) &=& \pi f^2 -\pi^2 f
\nonumber\\[-8pt]\\[-8pt]
&&{} +
\sum_{k=1}^{\infty} \bigl( \covardu{f \bigl(
\X[0]{i} \bigr)} {f \bigl(\X[k]{i} \bigr)} + \covardu{f \bigl(\X [1]{i} \bigr)} {f
\bigl( \X[k+1]{i} \bigr)} \bigr). \nonumber
\end{eqnarray}
For the kernels $P_i$ and $ Q_i$, $i\in\{0,1\}$, in the statement of
the theorem, let $\sequence{\R{i}}$, $i\in\{0,1\}$, be defined as in
Lemma~\ref{lemlem4}, which then implies that for all $\lambda\in
(0,1)$,
%
\begin{eqnarray}
\label{eqth4proof} && \sum_{k=1}^{\infty} \bigl(
\lambda^{k}\covardu{f \bigl(\X[0]{1} \bigr)} {f \bigl(\X[k]{1} \bigr)} +
\lambda^{k}\covardu{f \bigl(\X[1]{1} \bigr)} {f \bigl(\X[k+1]{1} \bigr)}
\bigr)
\nonumber\\[-8pt]\\[-8pt]
&&\qquad \leq
\sum_{k=1}^{\infty} \bigl(
\lambda^{k}\covardu{f \bigl(\X[0]{0} \bigr)} {f \bigl(\X[k]{0} \bigr)}+
\lambda^{k}\covardu{f \bigl(\X[1]{0} \bigr)} {f \bigl(\X[k+1]{0} \bigr)}
\bigr). \nonumber
\end{eqnarray}
We conclude the proof by letting $\lambda$ tend to one on each side of
the previous inequality. Under (\ref{eqassumpFuncThm}), we may, by the
dominated convergence theorem, interchange limits with summation, which
establishes inequality (\ref{eqth4proof}) also in the case
$\lambda
=1$. Combining this with (\ref{eqth4proof1}) completes the proof.
\end{pf*}

\section{Conclusion}In this paper, we have extended successfully the
theoretical framework proposed in \cite{peskunoptimum} and \cite
{tierneynote} as a means of comparing the asymptotic variance of
sample path averages for different Markov chains and, consequently, the
efficiency of different MCMC algorithms to the context of inhomogeneous
Markov chains evolving alternatingly according to two different Markov
transition kernels. It turned out that this configuration covers,
although not apparently, several popular MCMC algorithms such as
Randomized MCMC \cite{nichollscoupled}, Multiple-try Metropolis \cite
{liumultiple} and its generalization \cite{pandolfigeneralization},
and the pseudo-marginal algorithms \cite{andrieupseudo,andrieuconvergence}.
It should be remarked however that our results do not take possible
additional computational cost into consideration, which may be of
importance in practical applications. While these algorithms are
inapproachable for the standard tools provided in \cite{peskunoptimum}
and \cite{tierneynote}, our results allow, without heavy technical
developments, rigorous theoretical justifications advocating the use of
these algorithms. As illustrated by our novel
\textit{random refreshment} algorithm in the context of pseudo-marginal
algorithms, the results of the present paper can also be used for
designing new algorithms and improving, in terms of asymptotic
variance, existing ones.

\begin{appendix}
\section{Proofs of Propositions~\texorpdfstring{\lowercase{\protect\ref{propaltcondition}}}{9} and \texorpdfstring{\lowercase{\protect\ref{propinduces_rever}}}{13}}
\label{app}
\subsection{Proof of Proposition~\texorpdfstring{\protect\ref{propaltcondition}}{9}}
\label{appproofpropaltcondition}

First, set $\xi= P^n(x, \cdot) - \pi$; then by Jensen's inequality,
\[
\ensuremath{\Vert\xi\Vert_{V^{1/2}}} = |\xi|(\mathsf{X}) \frac
{|\xi
|(V^{1/2})}{|\xi|(\mathsf{X}
)}
\leq|\xi|( \mathsf{X}) \biggl( \frac{|\xi|(V)}{|\xi|(\mathsf{X})} \biggr)^{1/2} = |
\xi|^{1/2}( \mathsf{X}) \ensuremath{\Vert\xi\Vert_{V}}^{1/2},
\]
and since $|\xi|(\mathsf{X}) \leq2$,
%
\begin{equation}
\label{eqineq-Pn-V-half} \ensuremath{\bigl\Vert P^n(x, \cdot) - \pi\bigr\Vert
_{V^{1/2}}} \leq \bigl(2 C \rho^n V(x) \bigr)^{1/2}.
\end{equation}
Now, without loss of generality we may assume that $\pi f = 0$,
$|f|_{V^{1/2}} \leq1$, and $|Pf|_{V^{1/2}} \leq1$. Then applying
(\ref{eqineq-Pn-V-half}) yields for all $x \in\mathsf{X}$,
\[
\bigl|(PQ)^n f(x)\bigr| \leq \bigl(2 C \rho^n V(x)
\bigr)^{1/2}.
\]
Hence, for all $n \in\mathbb{N}$,
\begin{eqnarray*}
\bigl|\covardu{f \bigl(\X[0]{} \bigr)} {f \bigl(\X[2n]{} \bigr)}\bigr| &=& \bigl|\mathbb{E}
\bigl(f(X_0) (PQ)^n f(X_0) \bigr)\bigr|
\\
&\leq& \bigl(2 C \rho^n \bigr)^{1/2} \mathbb{E}
\bigl(\bigl|f(X_0)\bigr|V^{1/2}(X_0) \bigr) \leq \bigl(2
C \rho^{n} \bigr)^{1/2} \pi V.
\end{eqnarray*}
In the same way, for all $n \geq0$,
\[
\bigl|\covardu{f \bigl(\X[0]{} \bigr)} {f \bigl(\X[2n+1]{} \bigr)}\bigr|=\bigl|\mathbb{E}
\bigl(f(X_0) (PQ)^n Pf(X_0) \bigr)\bigr| \leq
\bigl(2 C \rho^{n} \bigr)^{1/2} \pi V.
\]
By applying successively the Cauchy--Schwarz and Jensen inequalities,
we obtain
\[
\mathbb{E} \bigl(\bigl|f(X_1)\bigr|QV^{1/2}(X_1) \bigr)
\leq \bigl[\mathbb{E} \bigl(f^2(X_1) \bigr) \mathbb{E}
\bigl(QV(X_1) \bigr) \bigr]^{1/2} \leq\pi V,
\]
where the last inequality follows from $f^2 \leq V$ and $\pi P=\pi
Q=\pi
$. This implies that for all $n \in\mathbb{N}^{\ast}$,
\begin{eqnarray*}
\bigl|\covardu{f \bigl(\X[1]{} \bigr)} {f \bigl(\X[2n]{} \bigr)}\bigr| &=& \bigl|\mathbb{E}
\bigl(f(X_1)Q(PQ)^{n-1} f(X_1) \bigr)\bigr|
\\
&\leq& \bigl(2 C \rho^{n-1} \bigr)^{1/2} \mathbb{E}
\bigl(\bigl|f(X_1)\bigr|QV^{1/2}(X_1) \bigr)
\\
&\leq& \bigl(2
C \rho^{n-1} \bigr)^{1/2} \pi V.
\end{eqnarray*}
In the same way, for all $n \in\mathbb{N}^{\ast}$ we have, using
that $|Pf(x)|
\leq V^{1/2}(x)$,
\[
\bigl|\covardu{f \bigl(\X[1]{} \bigr)} {f \bigl(\X[2n+1]{} \bigr)}\bigr| = \bigl|\mathbb{E}
\bigl(f(X_1)Q(PQ)^{n - 1} Pf(X_1) \bigr)\bigr| \leq
\bigl(2 C \rho^{n - 1} \bigr)^{1/2} \pi V.
\]
The statement of the proposition follows.

\subsection{Proof of Proposition~\texorpdfstring{\protect\ref{propinduces_rever}}{13}}
\label{appproofrevsystrefresh}

Let $K$ be the transition kernel of the Markov chain $\sequence{\Y
{2}}$, that is, for all $f \in\mathcal{F}(\mathcal{Y})$,
\begin{eqnarray*}
&& \int f \bigl(y' \bigr) K \bigl(y, \mathrm{d}y' \bigr)
\\
&&\qquad = f(y) \beta(y) + \int f \bigl(y' \bigr) R(y, \mathrm{d}u) S
\bigl(y, u; \mathrm{d}y' \bigr) T \bigl(y, u, y';
\mathrm{d}u' \bigr) \alpha \bigl(y, u, y ',
u' \bigr),
\end{eqnarray*}
where $\beta(y):=1 - \int R(y, \mathrm{d}u) S(y, u;
\mathrm{d}y') T(y, u, y'; \mathrm{d}u') \alpha(y, u, y
', u')$. Thus, establishing $\pi^{\ast}$-reversibility of $K$ amounts
to verifying, for all $f$ and $g$ in $\mathcal{F}(\mathcal{Y})$,
%
\begin{eqnarray}
\label{eqstargRev}
&& \int f(y)g \bigl(y' \bigr) \pi^{\ast}(
\mathrm{d}y) \int R(y,\mathrm{d}u) S \bigl(y,u;\mathrm{d}y' \bigr) T
\bigl(y, u,y';\mathrm{d}u' \bigr) \alpha \bigl(y,u,y
',u' \bigr)\nonumber
\\
&&\qquad = \int f(y)g \bigl(y' \bigr) \pi^{\ast} \bigl(
\mathrm{d}y' \bigr)
\\
&&\quad\qquad{}\times \int R \bigl(y',
\mathrm{d}u' \bigr) S \bigl( y',u';
\mathrm{d}y \bigr) T \bigl( y', u', y; \mathrm{d}u
\bigr) \alpha \bigl(y ',u',y,u \bigr). \nonumber
\end{eqnarray}
Indeed, by $\pi$-reversibility of $\{(\Y[k]{1},\Aux[k]{1}); k\in
\mathbb{N}
\}$ it holds, for all $\bar{f}$ and $\bar{g}$ in $\mathcal
{F}(\mathcal{Y} \varotimes \mathcal{U})$,
\begin{eqnarray*}
&& \iint\bar{f}(y, u) \bar{g} \bigl(y', u' \bigr) \pi(
\mathrm{d}y\times\mathrm{d}u) S \bigl(y, u; \mathrm{d} y' \bigr) T
\bigl(y, u, y'; \mathrm{d}u ' \bigr) \alpha \bigl(y, u,
y', u' \bigr)
\\
&&\qquad = \iint\bar{f}(y,u) \bar{g} \bigl(y', u' \bigr) \pi
\bigl(\mathrm{d}y' \times\mathrm{d}u' \bigr) S
\bigl(y',u';\mathrm{d}y \bigr)
\\
&&\hspace*{50pt}{}\times  T \bigl(
y',u',y;\mathrm{d}u \bigr) \alpha \bigl(
y',u',y,u \bigr),
\end{eqnarray*}
which establishes (\ref{eqstargRev}) by letting $\bar{f}(y, u) =
f(y)$ and $\bar{g}(y, u) = g(y)$. This completes the proof.

\section{Relation between Algorithm~\texorpdfstring{\lowercase{\protect\ref{algalg2}}}{2} and the \textup{r}-MCMC and GMTM
algorithms}\label{secappB}


%
\begin{algorithm}
\begin{algorithmic}[5]
\caption{r-MCMC \cite{nichollscoupled}} \label{algalgrMCMC}
\Require$\Y[k]{2} = y$:
\begin{longlist}[(iii)]
\item[(i)] draw $\tY\sim\check{R}(y, \cdot) \leadsto\hat{y}$,
\item[(ii)] draw $U\sim\check{S}(y, \hat{y}; \cdot) \leadsto u$,
\item[(iii)] set
%
\begin{equation}
\label{eqacceptrMCMC} \Y[k+1]{2} \gets%
\cases{
\hat{y}, &\quad w.pr. $\displaystyle \alpha^{(\mathrm{r})}(y, u, \hat{y})$
\vspace*{5pt}\cr
&\quad\qquad $\displaystyle:=1\wedge\frac{\pi^{\ast}(\hat{y})
\check{r}(\hat{y}, y) \check{s}(\hat{y}, y; f(u))}{\pi^{\ast}(y)\check{r}(y,\hat{y}) \check{s}(y, \hat{y}; u)} \bigg\vert\frac{\partial
f}{\partial u}(u) \bigg\vert$,
\vspace*{3pt}\cr
y, &\quad otherwise.}
\end{equation}
\end{longlist}
\end{algorithmic}
\end{algorithm}
%

\subsection{r-MCMC as a special case of Algorithm~\texorpdfstring{\protect\ref{algalg2}}{2}}\label{apprMCMC}
As proposed initially by~\cite{nichollscoupled}, the r-MCMC algorithm
generates a Markov chain $\sequence{\Y{2}}$ with transitions given by
Algorithm~\ref{algalgrMCMC} below. Denote\vspace*{2pt} by $\vert\frac{\partial
f}{\partial u}(u)\vert$ the Jacobian determinant of a
vector-valued transformation $f$.
In this\vspace*{2pt} algorithm, $f$ is any continuously differentiable involution on
$\mathsf{U}= \mathbb{R}^d$. In addition, $\check{R}$ and $\check
{S}$ are
instrumental kernels on $(\mathsf{Y}, \mathcal{Y})$ and $(\mathsf
{Y}^2,\mathcal{U})$,
respectively, having transition densities $\check{r}$ and $\check{s}$
with respect to some dominating measure and Lebesgue measure on
$\mathbb{R}
^d$, respectively.
%
\begin{prop} \label{apprmcmc}
The r-MCMC algorithm is a special case of Algorithm~\ref{algalg2}.
\end{prop}
\begin{pf}
Since $\tY$ and $U$, obtained in steps~(i) and (ii) of
Algorithm~\ref{algalgrMCMC}, are not drawn in the same order as in
Algorithm~\ref{algalg2}, we first derive the expression of the
corresponding kernels $R$ and $S$, that is,
\begin{eqnarray*}
R(y,\mathrm{d}u) &=& \biggl(\int\check{R}(y, \mathrm {d}
\hat{y}) \check{s}(y,\hat{y};u) \biggr) \lambda_d(\mathrm{d}u) = r(y,
u) \lambda_d(\mathrm{d}u),
\\
S(y, u; \mathrm{d}\hat{y}) & =& \frac{\check{R}(y, \mathrm{d}\hat{y})
\check
{s}(y, \hat{y}; u)}{\int\check{R}(y, \mathrm{d}\hat{y}) \check
{s}(y, \hat{y};u)},
\end{eqnarray*}
where $\lambda_d$ is Lebesgue measure on $\mathbb{R}^d$. Also note that
%
\begin{equation}
\label{eqtechnicos} R(y, \mathrm{d}u) S(y, u; \mathrm{d}\hat{y}) = \check {R}(y,
\mathrm{d} \hat{y}) \check{s}(y, \hat{y}; u) \lambda_d(\mathrm{d}u).
\end{equation}
Moreover, introduce another auxiliary variable $\tAuxi$ taking values
in $\mathsf{U}$ and being drawn according to $T(y, u, \hat{y};
\mathrm{d}
\hat{u})
= \delta_{f(u)}(\mathrm{d}\hat{u})$. Note that the kernel $T$ is not
dominated by a common nonnegative measure regardless the value of $u$;
still, following Remark~\ref{remgeneralradon}, the r-MCMC algorithm
may be covered by Algorithm~\ref{algalg2}, provided that the ratio in
the acceptance probability $\alpha^{(\mathrm{r})}(y, u, \hat{y})$
corresponds to the Radon--Nikodym derivative in Proposition~\ref
{propderiveeRadon} for
\[
K^{(\mathrm{r})}(y,u; \mathrm{d}\hat{y}\times\mathrm{d}\hat{u}) = S(y, u;
\mathrm{d}\hat{y}) T(y, u, \hat{y}; \mathrm{d}\hat{u}) = S(y, u; \mathrm{d}
\hat{y}) \delta _{f(u)}(\mathrm{d}\hat{u})
\]
and
\[
\pi^{(\mathrm{r})}(\mathrm{d}y\times\mathrm{d}u) = \pi^{\ast
}(
\mathrm{d}y) R(y, \mathrm{d} u).
\]
The proof is completed by applying Lemma~\ref{lemrMCMClem2} below.
\end{pf}

%
\begin{lem} \label{lemrMCMClem2}
The acceptance probability $\alpha^{(\mathrm{r})}$ in (\ref
{eqacceptrMCMC}) is equal to
%
\begin{equation}
\label{eqapp2} \alpha^{(\mathrm{r})}(y,u,\hat{y})=1\wedge\frac
{\mathrm{d}
\nu^{(\mathrm{r})}}{\mathrm{d}
\mu^{(\mathrm{r})}}(x,
\hat{x}),
\end{equation}
where $x:=(y,u)$, $\hat{x}:=(\hat{y},\hat{u})$, and
$\frac{\mathrm{d}
\nu^{(\mathrm{r})}}{\mathrm{d}\mu^{(\mathrm{r})}}$ denotes the
Radon--Nikodym
derivative between the measures $\nu^{(\mathrm{r})}$ and $\mu
^{(\mathrm{r})}$
defined by
\begin{eqnarray*}
\nu^{(\mathrm{r})}(\mathrm{d}x \times\mathrm{d}\hat
{x}) &:=& \pi^{(\mathrm{r})}(\mathrm{d}\hat{y}\times\mathrm{d}\hat{u})
K^{(\mathrm{r}
)}(\hat{y}, \hat{u}; \mathrm{d}y\times\mathrm{d}u),
\\
\mu^{(\mathrm{r})}(\mathrm{d}x \times\mathrm{d}\hat{x}) &:=& \pi ^{(\mathrm{r}
)}(
\mathrm{d}y \times\mathrm{d}u) K^{(\mathrm{r})}(y, u; \mathrm{d}\hat{y}\times
\mathrm{d}\hat{u}).
\end{eqnarray*}
\end{lem}
\begin{pf}
Write $\alpha^{(\mathrm{r})}(y, u, \hat{y}) = 1 \wedge\gamma
^{(\mathrm{r}
)}(y, u, \hat{y})$, where
\[
\gamma^{(\mathrm{r})}(y, u, \hat{y}):=\frac{\pi^{\ast}(\hat
{y})\check{r}(\hat{y}, y) \check{s}(\hat{y}, y; f(u))}{\pi^{\ast}(y) \check{r}(y,
\hat{y})\check
{s}(y, \hat{y}; u)} \bigg\vert
\frac{\partial f}{\partial u}(u) \bigg\vert.
\]
To show (\ref{eqapp2}), we will prove that for all bounded measurable
functions $G$ on $(\mathsf{Y}\times\mathsf{U})^2$ it holds that
\begin{eqnarray*}
\mathbb{E}_{\nu^{(\mathrm{r})}} \bigl[G(X, \tX) \bigr] &=& \int G(x,\hat{x})
\nu^{(\mathrm{r}
)}(\mathrm{d}x \times\mathrm{d}\hat{x})
\\
&=&\int G(x, \hat{x}) \gamma
^{(\mathrm{r}
)}(y, u, \hat{y}) \mu^{(\mathrm{r})}(\mathrm{d}x \times\mathrm{d}
\hat{x})
\end{eqnarray*}
[where $x = (y, u)$ and $\hat{x}= (\hat{y}, \hat{u})$]. Now, using the
change of variables $u=f(\hat{u})$, which is equivalent to $\hat{u}
=f(u)$ (since $f$ is an involution) and using the relation~(\ref{eqtechnicos}) we obtain
\begin{eqnarray*}
&& \mathbb{E}_{\nu^{(\mathrm{r})}}
\bigl[G^{(\mathrm{r})}(X, \tX) \bigr]
\\
&&\qquad = \int G^{(\mathrm{r})} \bigl(y,f(\hat{u}), \hat{y}, \hat{u} \bigr)
\pi^{\ast}(\mathrm{d}\hat{y}) r(\hat{y}, \hat{u}) S(\hat{y}, \hat{u};
\mathrm{d}y) \lambda _d(\mathrm{d} \hat{u})
\\
&&\qquad = \int G^{(\mathrm{r})} \bigl(y, u, \hat{y}, f(u) \bigr)
\\
&&\hspace*{42pt}{} \times \pi^{\ast
}(\mathrm{d} \hat{y}) r \bigl(\hat{y}, f(u) \bigr) S \bigl(\hat{y}, f(u);
\mathrm{d}y \bigr) \bigl|( \partial f/\partial u) (u)\bigr| \lambda_d(
\mathrm{d}u)
\\
&&\qquad = \int G^{(\mathrm{r})} \bigl(y, u, \hat{y}, f(u) \bigr) \frac{\pi
^{\ast}
(\hat{y}) \check
{r}(\hat{y}, y) \check{s}(\hat{y}, y; f(u))}{
\pi^{\ast}(y) \check{r}(y, \hat{y})\check{s}(y, \hat{y}; u)}
\bigg\vert\frac
{\partial f}{\partial u}(u) \bigg\vert
\\
&&\hspace*{9pt}\quad\qquad{} \times\pi^{\ast}(\mathrm{d}y) \check{R}(y,\mathrm {d}
\hat{y}) \check{S}( y, \hat{y}; \mathrm{d}u)
\\
&&\qquad = \int G^{(\mathrm{r})}(x, \hat{x}) \gamma^{(\mathrm{r})}(y, u, \hat{y}) \mu
^{(\mathrm{r})}(\mathrm{d}x \times\mathrm{d}\hat{x}),
\end{eqnarray*}
which completes the proof.
\end{pf}


\subsection{GMTM as a special case of Algorithm~\texorpdfstring{\protect\ref{algalg2}}{2}} \label{appMTM}
The GMTM algorithm proposed in \cite{pandolfigeneralization} generates
a Markov chain $\sequence{\Y{2}}$ with transitions given by\vadjust{\goodbreak}
Algorithm~\ref{algalgMTM} below.
\begin{algorithm}
\begin{algorithmic}[5]
\caption{GMTM \cite{pandolfigeneralization}} \label{algalgMTM}
\Require$\Y[k]{2} = y$:
\begin{longlist}[(iii)]
\item[(i)] draw $(V_1, \ldots, V_{n}) \sim_{\mathrm{i.i.d.}}
\check
{R}(y, \cdot) \leadsto(v_1, \ldots, v_n)$,
\item[(ii)] let $J$ take the value $j \in\{1, 2, \ldots, n \}$ w.pr.
$\omega
(y,v_j) / \sum_{\ell= 1}^n \omega(y,v_\ell)$,
\item[(iii)] let $\hat{y}\gets v_j$,
\item[(iv)] draw $(\tAuxiv_1, \ldots, \tAuxiv_{n - 1}) \sim_{\mathrm{i.i.d.}}
\check{R}(\hat{y}, \cdot) \leadsto(\hat{v}_1, \ldots, \hat{v}_{n
- 1})$,
\item[(v)] let $\hat{v}_n \gets y$,
\item[(vi)] let
%
\begin{equation}
\label{eqacceptMTM} \Y[k+1]{2} \gets%
\cases{
\hat{y}, &\quad with probability $\displaystyle \alpha^{(\mathrm{m})}(y, v, \hat{y}, \hat{v})$
\vspace*{5pt}\cr
&\quad\qquad $\displaystyle :=  1 \wedge\frac{\pi^{\ast}(\hat{y}) \check{r}(\hat{y}, y) \omega(\hat{y}, y) \sum_{k = 1}^n \omega(y,v_k)}{\pi^{\ast} (y) \check{r}(y, \hat{y}) \omega(y, \hat{y}) \sum_{k = 1}^n
\omega(\hat{y},\hat{v}_k)}$,
\vspace*{3pt}\cr
y, &\quad otherwise.}
\end{equation}
\end{longlist}
\end{algorithmic}
\end{algorithm}
In Algorithm~\ref{algalgMTM}, the auxiliary variables $V_1,
\ldots, V_n$ are defined on $\mathsf{Y}$ and for all $y\in\mathsf{Y}$ and $(v
_1, \ldots, v_n) \in\mathsf{Y}^n$, $\{\omega(y, v_k) /\break  \sum_{\ell
= 1}^n \omega(y, v_\ell) \}_{k = 1}^n$ are sample weights.
Moreover, $\check{R}$ is an instrumental kernel defined on $(\mathsf{Y},\mathcal{Y}
)$ having the transition density $\check{r}$ with respect to some
dominating measure on $(\mathsf{Y},\mathcal{Y})$.
%
\begin{prop} \label{lemMTM}
The GMTM algorithm is a special case of Algorithm~\ref{algalg2}.
\end{prop}
\begin{pf}
Denoting by $V_1, \ldots, V_n$ the random variables
generated in step~(i) in Algorithm~\ref{algalgMTM},
the proposed candidate $\tY$ is obtained as $V_J$, where $J$ is
generated in step~(ii).
Let $U= V_{-J}$, where
\[
v_{-j}:=(v_1, \ldots, v_{j - 1},
v_{j + 1}, \ldots, v_n).
\]
To obtain the joint distribution of $(\tY,U)$ conditionally on
$\Y
[k]{2}$, write for any bounded measurable function $G$ on $\mathsf{Y}^n$,
\begin{eqnarray*}
&& \mathbb{E} \bigl[G(\tY,U) \mid\Y[k]{2} = y
\bigr]
\\
&&\qquad = \sum_{j = 1}^n \mathbb{E} \bigl[G(
V_j, V_{-j}) \mathbh{1}_{J = j} \mid \Y[k]{2} =
y \bigr]
\\
&&\qquad = \idotsint\check{R}(y, \mathrm{d}\hat{y}) \prod_{k = 1}^{n - 1}
\check{R}(y, \mathrm{d}u_k) \frac{ n \omega(y, \hat{y})}{\sum_{\ell
= 1 }^{n - 1}
\omega(y, u_\ell) + \omega(y, \hat{y})}G(\hat{y}, u)
\\
&&\qquad = \idotsint R(y, \mathrm{d}u) S(y, u; \mathrm{d}\hat{y}) G(\hat {y}, u),
\end{eqnarray*}
where we introduced the kernels
%
\begin{eqnarray}
R(y,\mathrm{d}u) &:=& n \prod_{k = 1}^{n - 1}
\check{R}(y, \mathrm{d}u_k) \int\frac{\check{R}(y, \mathrm{d}\hat{y})
\omega(y, \hat{y})}{\sum_{\ell
= 1}^{n - 1} \omega(y, u_\ell) + \omega(y, \hat{y})},  \label
{eqMTMR}
\\
\qquad S(y, u; \mathrm{d}\hat{y}) &:=& \frac{\check{R}(y, \mathrm{d}
\hat{y})
\omega(y, \hat{y})}{\sum_{\ell= 1 }^{n - 1} \omega(y, u_\ell) +
\omega(y, \hat{y})} \Big/ {\int
\frac{\check{R}(y, \mathrm{d}\hat{y}) \omega
(y, \hat{y})}{\sum_{\ell= 1 }^{n - 1} \omega(y, u_\ell) +
\omega(y,
\hat{y})}}.  \label{eqMTMS}
\end{eqnarray}
Now, set $\tAuxi= (\tAuxiv_1,\ldots,\tAuxiv_{n-1})$ where the $\tAuxiv_i$'s
are sampled in step~(iv). The distribution of $\tAuxi$ conditionally on
$(\Y[k]{2}, U, \tY) = (y, u, \hat{y})$ is given by
%
\begin{equation}
\label{eqMTMT} T(y, u, \hat{y}; \mathrm{d}\hat{u}) = \prod
_{k = 1}^{n - 1} \check{R}(\hat{y}, \mathrm{d}
\hat{u}_k).
\end{equation}
If $\check{R}$ is dominated by a nonnegative measure, then (\ref
{eqMTMR}), (\ref{eqMTMS}) and (\ref{eqMTMT}) show that the kernels
$R$, $S$ and $T$ are dominated as well. Denoting by $r$, $s$ and $t$
the corresponding transition densities, it can be checked readily that
\begin{eqnarray*}
&& \frac{\pi^{\ast}(\hat{y}) r(\hat{y}, \hat{u}) s(\hat{y}, \hat
{u}; y
) t(\hat{y}, \hat{u}, y;
u)}{\pi^{\ast}(y) r(y, u) s(y, u; \hat{y}) t(y, u, \hat{y};
\hat{u})}
\\
&&\qquad = \frac{\pi^{\ast}(\hat{y}) \check{r}(\hat{y}, y) \omega(\hat{y},
y)(\sum_{k = 1}^{n
- 1} \omega(y, u_k) + \omega(y, \hat{y}))}{\pi^{\ast}(y) \check
{r}(y,
\hat{y})\omega(y, \hat{y})(\sum_{k = 1}^{n - 1} \omega(\hat{y},
\hat{u}_k) + \omega
(\hat{y}, y))},
\end{eqnarray*}
so that $\alpha^{({m})}$ defined in (\ref{eqacceptMTM}) corresponds
to the acceptance probability $\alpha$ defined in~(\ref{eqacceptMetropolis}) with these particular choices of $r$, $s$ and
$t$. Consequently, the GMTM algorithm is a special case of
Algorithm~\ref{algalg2}.
\end{pf}
Note that in the previous proof, we have chosen the auxiliary variable
$U$ as the vector of rejected candidates after step~(ii). Another
natural idea would consist in choosing $U= (V_1, \ldots,
V_n)$, where the $V_i$s are obtained in step~(i); however,
since $\tY$ belongs to this set of candidates, the model would then not
be dominated, which would make the proof more intricate.
\end{appendix}


\section*{Acknowledgements}
We thank the anonymous referees for insightful comments that improved
significantly the presentation of the paper. A special thanks goes to
the referee who provided the two counterexamples in Remarks~\ref{remcounterexsummability}~and~\ref{remcounterexlemma}, as
well as the possible application of our methodology to the ABC context
in Example~\ref{exrrABC}.


%

\printaddresses

\end{document}